\documentclass[epj,fleqn]{svjour}
\journalname{The EPJ Plus}
\usepackage[english]{babel}
\usepackage{amsmath}
\usepackage{amssymb}
\usepackage{upgreek}
\usepackage[T1]{fontenc}
\usepackage[x11names]{xcolor}
\usepackage{empheq}
\usepackage{color}
\allowdisplaybreaks
\usepackage{mathtools}
\usepackage{eucal}
\usepackage{graphicx}
\usepackage{float}
\usepackage{bookmark}
\usepackage{breakurl}
\usepackage{epsfig}
\usepackage{bm}
\usepackage[utf8x,latin1]{inputenc} 
\usepackage{emerald}
\usepackage{slantsc}
\usepackage{array}
\usepackage[sort&compress,numbers]{natbib}

\def\be{\begin{equation}}
\def\ee{\end{equation}}
\def\ba{\begin{eqnarray}}
\def\ea{\end{eqnarray}}
\def\bsu{\begin{subequations}}
\def\esu{\end{subequations}}

\def\R{{\mathcal R}}

\def\B{{\rm B}}

\def\R{{\rm R}}

\def\a{\alpha}
\def\b{\beta}
\def\g{\gamma}     
\def\G{\Gamma}
\def\d{\delta}

\def\l{\lambda}
\def\m{\mu}
\def\n{\nu}

\def\r{\rho}
\def\s{\sigma}
\def\t{\tau}

\def\rmg{{\rm g}}
\def\rmh{{\rm h}}

\def\la{\label}
\def\pd{\partial}

\begin{document}
\onecolumn
\title{Covariant Equations of Motion Beyond the Spin-Dipole Particle Approximation}
\author{Sergei M. Kopeikin 
	}                     
%
%
\institute{Department of Physics \& Astronomy, University of Missouri, 322 Physics Bldg., Columbia, Missouri 65211, USA\\
\email{kopeikins@missouri.edu}}
\date{Received: date / Revised version: date}
\abstract{The present paper studies the post-Newtonian dynamics of $N$-body problem in general relativity. We derive covariant equations of translational and rotational motion of $N$ extended bodies having arbitrary distribution of mass and velocity of matter by employing the set of global and local coordinate charts on curved spacetime manifold $M$ of $N$-body system along with the mathematical apparatus of the Cartesian symmetric trace-free tensors and Blanchet-Damour multipole formalism. We separate the self-field effects of the bodies from the external gravitational environment and construct the effective background spacetime manifold by making use of the asymptotic matching technique. We make worldline of the center of mass of each body identical with that of the origin of the body-adapted local coordinates by the appropriate choice of the dipole moments. The covariant equations of motion are obtained on the background manifold $\bar M$ by applying the Einstein principle of equivalence and the Fermi-Walker law of transportation of the linear momentum and spin of each body. Our approach significantly extends the Mathisson-Papapetrou-Dixon covariant equations of motion beyond the spin-dipole particle approximation by accounting for the entire infinite set of the internal multipoles of the bodies which are gravitationally coupled with the curvature tensor of the background manifold $\bar M$ and its covariant derivatives. The results of our study can be used for much more accurate prediction of orbital dynamics of extended bodies in inspiraling binary systems and construction of templates of gravitational waves at the merger stage when the strong gravitational interaction between the higher-order multipoles of the bodies play a dominant role. The covariant theory of the post-Newtonian equations of motion beyond the spin-dipole approximation is a solid foundation for future improvements in long-term accuracy of relativistic celestial ephemerides of the solar system bodies.     
\PACS{{04.20.Cv}{fundamental problems and general formalism}   \and
      {04.25.-g}{approximation methods; equations of motion}  \and
      {04.25.Nx}{post-Newtonian approximation; perturbation theory}  \and
      {95.10.Ce}{N-body problem;celestial mechanics}
      }  
} 
\maketitle
\tableofcontents

\section{Introduction}\la{intro}

Post-Newtonian dynamics of an isolated gravitating system consisting of $N$ extended bodies moving on curved spacetime manifold $M$ is known in literature as {\it relativistic celestial mechanics} -- the term coined by Victor Brumberg \citep{vab,brum}. Mathematical properties of the manifold $M$ are fully determined in general relativity by the metric tensor $g_{\a\b}$ which is found by solving Einstein's field equations. General-relativistic celestial mechanics admits a minimal number of fundamental constants characterizing geometry of spacetime -- the universal gravitational constant $G$ and the "speed of light" $c$. Fundamental speed, $c$ is called the "speed of light" for historical reason for it was first measured in electrodynamics as the ratio of the electromagnetic and electrostatic units of charge, $c=\left(\epsilon_0\mu_0\right)^{-1/2}$. However, $c$ enters all four sectors of fundamental interactions and has different physical meaning depending on the type of interaction \citep{uzan_2005AmJPh}. In particular, the fundamental speed $c$ in the gravity sector of general relativity is the speed of propagation of gravity -- the "speed of gravity" \citep{Kopeikin_2004CQGra}. Its numerical value has been experimentally measured for the first time by \citet{Fomalont_2003ApJ} by using VLBI observation of gravitational bending of radio-waves in time-dependent gravitational field of moving Jupiter \citep{Kopeikin_2001ApJ,Kopeikin_2006FoPh}. Fomalont-Kopeikin measurement is sensitive to the first time derivatives of gravitational field and should not be confused with the direct measurement of the speed of propagation of gravitational waves which is sensitive to the second time derivatives of the metric tensor \citep{Kopeikin_2004CQGra,2007GReGr..39.1583K} or with the measurement of the speed of light from the quasar as some researchers erroneously suggested due to misinterpretation of $c$ entering the gravity sector of the theory \citep{kop_2005CQGra22,2006IJMPD..15..305K}.

Post-Newtonian celestial mechanics deals with an isolated gravitating ${N}$-body system which theoretical concept cannot be fully understood without careful study of three aspects -- asymptotic structure of spacetime, approximation methods and equations of motion \citep{Ehlers_1980NYASA,frau_2004LRR}.
In what follows, we adopt that spacetime is asymptotically-flat at infinity and the post-Newtonian (PN) approximations can be applied for solving the field equations. Strictly speaking, this assumption is not valid as our physical universe is described by Friedmann-Lema\^{i}tre-Robertson-Walker (FLRW) metric which is conformally-flat at infinity. Relativistic dynamics of extended bodies in FLRW universe requires development of the post-Friedmannian approximations for solving field equations in case of an isolated gravitating system placed on the FLRW spacetime manifold. The post-Friedmannian approximation method is more fundamental than the PN approximations and includes additional small parameter that is the ratio of the characteristic length of the isolated gravitating system to the Hubble radius of the universe. Rigorous mathematical approach for doing the post-Friedmannian approximations is based on the theory of Lagrangian perturbations of pseudo-Riemannian manifolds \citep{Petrov_2017book} and it has been worked out in a series of our papers \citep{Kopeikin_2013PhRvD,Kopeikin_2014AnPhy}. Relativistic celestial mechanics of an isolated gravitating systems in cosmology leads to a number of interesting predictions \citep{Kopeikin_2012PhRvD,galkop_2016PhRvD}. 

Equations of motion of ${N}$-body system describe the time evolution of a set of independent variables in the configuration space of the system. These variables are volume integrals from the continuous distribution of matter introduced by \citet{bld} and known as  mass and spin (or current) multipoles of gravitational field. Among them, mass-monopole, mass-dipole and spin-dipole of each body play a primary role in the description of translational and rotational degrees of freedom. Higher-order multipoles of each body couples with the external gravitational field of other bodies of the isolated system and perturbs the evolution of the lower-order multipoles of the body in the configuration space.  Equations of motion are subdivided into three main categories corresponding to various degrees of freedom of the configuration variables of the $N$-body system \citep{fockbook}. They are:
\begin{enumerate}
\item[1)] translational equations of motion of the linear momentum and the center of mass of each body,
\item[2)] rotational equations of motion\index{equations of motion!rotational} of the intrinsic angular momentum (spin) of each body,
\item[3)] evolutionary equations of the higher-order (quadrupole, octupole, etc.) multipoles of each body.
\end{enumerate}
Translational and rotational equations of motion in general relativity are sufficient to describe the dynamics of the spin-dipole massive particles which are assumed to be physically equivalent to spherically-symmetric and rigidly-rotating bodies. Deeper understanding of celestial dynamics of arbitrary-structured extended bodies requires derivation of the evolutionary equations of the higher-order multipoles. Usually, a simplifying assumption of the rigid rotation about the center of mass of each body is used for this purpose \citep{fockbook,brum,spyrou_1975ApJ,arminjon_2005PhRvD,Racine_2006CQG}. However, this assumption works well until one can neglect the tidal deformation of the body caused by the presence of other bodies in the $N$-body system and, certainly, cannot be applied at the latest stages of binary orbital evolution before merger. It is worth noticing that  some authors refer to the translational and rotational equations of the linear momentum and spin of the bodies as to the laws of motion and precession \citep{Havas_1962PhRv,Ehlers_1980NYASA,th_1985,Zhang_1985PhRvD} relegating the term {\it equations of motion} to the center of mass and angular velocity of rotation of the bodies. We do not follow this terminology in the present paper.

The most works on the equations of motion of massive bodies have been done in particular coordinates from which the most popular are the ADM and harmonic coordinates \citep{memmesh_2005PhRvD,hergt_2008PhRvD,schaefer_2011mmgr} \footnote{The ADM and harmonic coordinate charts are in general different structures but they can coincide under certain circumstances \citep{kopeikin_1999PhRvD}.}. However, the coordinate description of relativistic dynamics of ${N}$-body system must have a universal physical meaning and predict the same dynamical effects irrespective of the choice of coordinates on spacetime manifold $M$. The best way to eliminate the appearance of possible spurious coordinate-dependent effects would be a derivation of covariant equations of motion based entirely on the covariant definition of the configuration variables. To this end \citet{mathisson_2010GReGr_1,mathisson_2010GReGr_2}, \citet{Papapetrou23101951,pap1} and, especially, \citet{dixon_1970_1,dixon_1970_2,dixon_1974_3,dixon_1973GReGr,dixon_1979,dixon_2008,Dixon2015} had published a series of programmatic papers suggesting constructive steps toward the development of such fully-covariant algorithm of derivation of the covariant equations of motion (see also \citep{Taub_1965,madore_1969}) known as Mathisson's {\it variational dynamics} or the Mathisson-Papapetrou-Dixon (MPD) formalism \citep{dixon_2008,Dixon2015}. The MPD formalism pursued an ambitious goal to make it applicable to arbitrary metric-based theory of gravity but this created a number of hurdles that slowed down developing the covariant dynamics of extended bodies. Nonetheless, theoretical work on various aspects of the MPD theory has never stopped \citep{ehlers_1977GReGr,schattner_1979GReGr,Ohashi_2003PRD,steinhoff_2010PhRvD,dirk_2013PhLA,Obukhov_Puetzfeld2014,dirk_obukhov2014,dixon_2008}.  

In order to link the covariant MPD formalism to the coordinate-based derivations of equations of motion of extended bodies it should be extended to include a recipe of construction of the effective background manifold $\bar M$. Moreover, the Dixon multipoles \citep{dixon_1973GReGr,dixon_1979} have to be compared to the Blanchet-Damour multipoles of gravitational field. To find out these missing elements of the MPD formalism we tackle the problem of the covariant formulation of the equations of motion in a particular gauge associated with the class of harmonic coordinates. We build the effective background manifold $\bar M$ as a regular solution of the Einstein field equations and apply the Einstein equivalence principle for deriving covariant equations of motion by mapping the Blanchet-Damour multipoles to 4-dimensional form which can be compared with the covariant form of the Dixon multipoles. This procedure has been consistently developed and justified by \citet{th_1985}. 

Dynamics of an isolated gravitating system consisting of ${N}$ extended bodies is naturally split in two parts -- the relative motion of the bodies with respect to each other and the temporal evolution of the Blanchet-Damour multipoles of each body transported along worldline ${\cal Z}$ of body's center of mass. It suggests separation of the problem of motion in two parts: external and internal \citep{fockbook,Damour_1987book}. The external problem deals with the derivation of translational equations of motion of the body-adapted local coordinates. Solution of the internal problem provides us with definition of the Blanchet-Damour multipoles and local equations of motion of the center of mass of body with respect to the body-adapted local coordinates. Besides, the internal problem also gives us the evolutionary equations of the body's multipoles including rotational equations for spin. Solution of the external problem is rendered in a single global coordinate chart covering the entire spacetime manifold $M$. Solution of the internal problem is executed separately for each body in the body-adapted local coordinates. There are ${N}$ local coordinate charts -- one for each body -- making the atlas of the spacetime manifold $M$. Mathematical construction of the global and local coordinates is achieved through the solutions of the Einstein field equations. The coordinate-based approach to solving the problem of motion provides the most effective way for unambiguous separation of internal and external degrees of freedom of configuration variables by matching asymptotic expansions of the metric tensors in the local and global coordinates. Matching allows to find out the structure of the coordinate transformations between the local and global charts of the atlas of manifold $M$ and to build the effective background manifold $\bar M$ that is used for prolongation of equations of motion from the local chart to covariant form which is compared with Dixon's covariant equations of motion.  

The global coordinate chart is introduced for describing the orbital dynamics of the body's center of mass. It is not unique but defined up to the group of diffeomorphisms which leaves spacetime asymptotically-flat at null infinity. This is the Bondi-Metzner-Sachs (BMS) group \citep{winicour_2016,Schmidt_1975GReGr} that includes the Poincar\'e transformations as a sub-group.  It means that we can always introduce a non-rotating global coordinate chart with the origin located at the center of mass of the $N$-body system such that at infinity: (1) the metric tensor approaches the Minkowski metric, $\eta_{\alpha\beta}$\index{Minkowski metric}, and (2) the global coordinates smoothly match the inertial (Lorentzian) coordinates of the Minkowski spacetime. The global coordinate chart is not sufficient for solving the problem of motion of extended bodies as it is not adequately adapted for the description of internal structure and motion of matter inside each body in the isolated ${N}$-body system. This description is done more naturally in a local coordinate chart attached to each gravitating body. Properly chosen local coordinates exclude a number of spurious effects appearing in the global coordinates but having no physical relation to the intrinsic motion of body's matter \citep{Kopejkin_1988CeMec}. The body-adapted local coordinates replicate the inertial Lorentzian coordinates only in a limited domain of spacetime manifold $M$ inside a world tube around the body under consideration. Thus, a complete coordinate-based solution of the external and internal problems of celestial mechanics requires introduction of ${N}+1$ coordinate charts -- one global and ${N}$ local ones \citep{iau2000,Battista_2017IJMPA}. It agrees with the topological structure of spacetime defined by a set of the overlapping coordinate charts making the atlas of spacetime manifold $M$ \citep{dfn}. The equations of motion of the bodies are intimately connected to the differential structure of the manifold $M$ characterized by the metric tensor and its derivatives. It means that the functional forms of the metric tensor in the local and global coordinates must be diffeomorphically equivalent. The principle of covariance is naturally satisfied by the law of transformation from the global to local coordinates. 

The brief content of our study is as follows. Next section \ref{notat} summarizes the main concepts and notations used in the present paper. Section \ref{tv3ef} summarizes the basic elements of the MPD formalism and presents covariant equations     
of motion derived by \citet{dixon_1979}. Atlas of spacetime manifold $M$ in $N$-body problem is explained in section \ref{qw231z}. The procedure of matching of 
the asymptotic expansions of the metric tensor in the global and local coordinate charts is described in section \ref{nrv27x}. It defines the 
transition functions between the coordinate charts and yields the equation of motion for worldline ${\cal W}$ of the origin of the local chart adapted to body B. Section \ref{km2va} provides the reader with the definitions of the Blanchet-Damour internal multipoles of body B. It also defines gravitoelectric and gravitomagnetic external multipoles of the body. Section \ref{mop3vx4} derives equations of motion for linear momentum and spin of body B in the local coordinates of the body and fixes the center of mass of body B at the origin of its own local coordinates. This makes worldline ${\cal W}$ of the origin of the local coordinates identical with the worldline ${\cal Z}$ of the center of mass of the body.  The effective background manifold $\bar M$ of body B and the background metric $\bar g_{\a\b}$ of this manifold are constructed in section \ref{po3v6} for each extended body. The background manifold $\bar M$ is the arena for derivation of the covariant equations of motion of the extended bodies. Section \ref{o8b4m1} proves that the center of mass of body B moves along perturbed time-like geodesic on the background manifold $\bar M$. The perturbation is caused by the gravitational interaction between the internal and external multipoles of the body. Section \ref{n4v7a9} extends 3-dimensional internal and external multipoles to 4-dimensional spacetime. Section \ref{nxv34xd} converts the equations of motion derived in section \ref{mop3vx4} to a covariant form. Section \ref{appndxon} establishes mathematical correspondence between the covariant form of the Dixon and Blanchet-Damour multipole moments. Section \ref{appendixB} compares the Dixon covariant equations of translational and rotational motion of extended bodies with our covariant equations of motion from section \ref{nxv34xd}. 

\section{Notations and Conventions}\label{notat}

We work in the geometrized system of units such that $G=c=1$. 

We consider an isolated gravitating system consisting of $N$ extended bodies in the framework of general relativity. The bodies are indexed by the capital Roman letters B and C each taking values from 1 to $N$. The bodies have generic distribution of mass density $\rho$, internal energy density $\Pi$, pressure $p$, and velocity of matter which depend on time. We exclude exchange of matter between the bodies and processes of nuclear transmutation of matter particles.  

Describing motion of $N$-body system requires introduction of one global coordinate chart, $x^\a=(t,x^i)$, covering the entire spacetime manifold and $N$ local coordinate charts, $w^\a=(u,w^i)$, adapted to each body B of the $N$-body system. The metric outside body B is parametrized by two infinite sets of configuration parameters which are called the {\it internal} and {\it external} multipoles. The multipoles are purely spatial, 3-dimensional, symmetric trace-free (STF) Cartesian tensors \citep{Pirani1964,thor,bld1986} residing on the hypersurface of constant coordinate time $u$ passing through the origin of the local coordinate chart, $w^\a$. The internal multipoles characterize  gravitational field and internal structure of body B and are of two types -- the mass multipoles ${\cal M}^L$, and the spin multipoles ${\cal S}^L$ where the multi-index $L=i_1i_2\ldots i_l$ consists of a set of spatial indices with $l$ denoting the rank of the STF tensor. There are also two types of external multipoles -- the gravitoelectric multipoles ${\cal Q}_L$, and the gravitomagnetic multipoles ${\cal C}_L$. The external multipoles with rank $l\ge 2$ characterize tidal gravitational field in the neighborhood of body B produced by other (external) bodies residing outside body B. Gravitoelectric dipole ${\cal Q}_i$ describes local acceleration of the origin of the local coordinates adapted to body B. Gravitomagnetic dipole ${\cal C}_i$ is the angular velocity of rotation of the spatial axes of the local coordinates. In what follows we set ${\cal C}_i=0$. The above-mentioned multipoles are called {\it canonical} as they are intimately related to two degrees of freedom of vacuum gravitational field. The overall theory also admits the appearance of {\it non-canonical} STF multipoles at the intermediate steps in the course of derivation of the equations of motion. These multipoles are gauge-dependent and can be eliminated from the equations of motion by using the residual gauge freedom and by adjusting the values of the linear and angular momenta of the bodies.

Definitions of the {\it canonical} STF multipoles must be consistent with the differential structure of spacetime manifold $M$ determined by the solutions of Einstein's field equations in the global and local coordinate charts. The consistency is achieved by applying the method of asymptotic matching of the external and internal solutions of the field equations that allows us to express the external multipoles, ${\cal Q}_L$ and ${\cal C}_L$, in terms of the internal multipoles, ${\cal M}^L$ and ${\cal S}^L$. The internal multipoles of an extended body B are defined by the integrals taken over body's volume from the internal distribution of mass-energy and stresses inside the body and the energy density of the tidal gravitational field produced by the external bodies.  

There are two important reference worldlines associated with motion of each body B -- a worldline ${\cal W}$ of the origin of the body-adapted, local coordinates $w^\a$, and a worldline ${\cal Z}$ of the center of mass of the body. Equations of motion of the origin of the local coordinates are obtained by matching of the internal and external solutions of Einstein's equations for the metric tensor. Equations of motion of the center of mass of body B are derived by integrating the macroscopic post-Newtonian equations of motion of matter which are consequence of the local law of conservation of the stress-energy tensor. The center of mass of each body is defined by the condition of vanishing of the internal mass dipole of the body in the multipolar expansion of the metric tensor, ${\cal M}^i=0$. This definition imposes a constraint on the local acceleration ${\cal Q}_i$ that makes worldline ${\cal W}$ coinciding with ${\cal Z}$. It also eliminates the {\it non-canonical} types of STF multipoles of gravitational field.       

Primary mathematical symbols and notations used in the present paper are as follows:
\begin{itemize}
\item the small Greek letters $\a,\b,\g,\ldots$ denote spacetime indices of tensors and run through values $0,1,2,3$, 
\item the small Roman indices $i,j,k,\ldots$ denote spatial tensor indices and take values $1,2,3$,
\item the capital Roman letters $L,K$ denote spatial tensor multi-indices, for example, $L\equiv \{i_1i_2\ldots i_l\}$, $K-1\equiv \{i_1i_2\ldots i_{k-1}\}$, etc.,  
\item the Einstein summation rule is applied for repeated (dummy) indices and multi-indices,
\item $\delta_{ij}$ is the Kronecker symbol,  
\item $\varepsilon_{ijk}=\varepsilon^{ijk}$ is 3-dimensional anti-symmetric symbol of Levi-Civita, 
\item $E_{\a\b\g\d}$ is 4-dimensional, fully anti-symmetric Levi-Civita symbol,
\item $\eta_{\a\b}={\rm diag}\{-1,+1,+1,+1\}$ is the Minkowski metric,
\item $g_{\a\b}$ is the metric of spacetime manifold in the global coordinates,
\item ${\rm g}_{\a\b}$ is the metric of spacetime manifold in the local coordinates adapted to body B,
\item $h_{\a\b}$ is the metric perturbation of the Minkowski spacetime in the global coordinates,
\item ${\rm h}_{\a\b}$ is the metric perturbation of the Minkowski spacetime in the local coordinates adapted to body B,
\item $\pd_\a=\pd/\pd x^\a$ is a partial derivative with respect to coordinate $x^\a$,
\item shorthand notations for multi-index partial derivatives with respect to coordinates $x^\a$ are: $\partial_L\equiv\partial_{i_1\ldots i_l}=\pd_{i_1}\pd_{i_2}...\pd_{i_l}$, $\partial_{L-1}\equiv\partial_{i_1\ldots i_{l-1}}$, $\partial_{pL-1}\equiv\partial_{pi_1...i_{l-1}}$, etc.,
\item $\nabla$ denotes a covariant derivative,
\item tensor indices of geometric objects on spacetime manifold are raised and lowered with the full metric $g_{\a\b}$,
\item tensor indices of geometric objects on the effective background manifold are raised and lowered with the background metric $\bar g_{\a\b}$,
\item tensor indices of the metric tensor perturbation $h_{\a\b}$ are raised and lowered with the Minkowski metric $\eta_{\a\b}$,
\item the spatial (Roman) indices of geometric objects are raised and lowered with the Kronecker symbol $\delta^{ij}$,
\item the ordinary factorial is $l!=l(l-1)(l-2)...2\cdot 1$, 
\item the round parentheses around a group of tensor indices denote a full symmetrization, \\
$T_{(\a_1\a_2...\a_l)}=\displaystyle\frac1{l!}\sum_{\sigma\in S} T_{\sigma(\a_1)\sigma(\a_2)...\sigma(\a_l)}$,
where $\sigma$ is a permutation of the set $S=\{\a_1,\a_2,...,\a_l\}$,
\item the square parentheses around a pair of tensor indices denote anti-symmetrization, for example,\\
$T^{[\a\b]\g}=\displaystyle\frac12\left(T^{\a\b\g}-T^{\b\a\g}\right)$,
\item the angular brackets around tensor indices denote a symmetric trace-free (STF) projection of tensor $T_L=T_{i_1i_2...i_l}$. The STF projection $T_{<L>}$ of tensor $T_L$ is constructed from its symmetric part, $T_{(L)}\equiv T_{(i_1i_2...i_l)}$
by subtracting all the permissible traces. This makes $T_{<L>}$ fully-symmetric and trace-free on all pairs of indices. The general formula for the STF projection is \citep{Pirani1964,thor,bld1986}
\be\label{stfformula}
T_{<L>}\equiv\sum_{n=0}^{[l/2]}\frac{(-1)^n}{2^nn!}\frac{l!}{(l-2n)!}\frac{(2l-2n-1)!!}{(2l-1)!!}\delta_{(i_1i_2...}\delta_{i_{2n-1}i_{2n}}T_{(i_{2n+1}...i_l)j_1j_1...j_nj_n)}\;,
\ee
where $[l/2]$ is the largest integer less than or equal to $l/2$.
\item the STF partial derivative is denoted by the angular parentheses embracing the STF indices, for example, $\pd_{<L>}\equiv\pd_{<i_1i_2...i_l>}$ or $\pd_{<K>}\equiv \pd_{<i_1i_2...i_k>}$\;,
\item the Christoffel symbols on spacetime manifold are: $\Gamma^\a_{\b\g}=\frac12 g^{\a\s}\left(\pd_\b g_{\g\s}+\pd_\g g_{\b\s}-\pd_\s g_{\b\g}\right)$\;,
\item the Riemann tensor on spacetime manifold $M$ is defined by convention (\citep[Equation 6.6.2]{weinberg_book1972} with an opposite sign)
\be\label{kk33cc25}
R_{\a\b\m\n}=\frac12\left(\pd_{\a\n} g_{\b\m}+\pd_{\b\m} g_{\a\n}-\pd_{\b\n} g_{\a\m}-\pd_{\a\m} g_{\b\n}\right)+ g_{\r\s}\left(\G^\r_{\a\n}\G^\s_{\b\m}-\G^\r_{\a\m}\G^\s_{\b\n}\right)\;.
\ee 
\end{itemize}
Other notations will be introduced and explained in the main text of the paper as they appear.

\section{Dixon's Theory of Equations of Motion}\la{tv3ef}

The goal to build a covariant post-Newtonian theory of motion of extended bodies and to find out the relativistic corrections to the equations of motion of a point-like particle which account for {\it all} multipoles characterizing the interior structure of the extended bodies was put forward by Mathisson\index{Mathisson} \citep{mathisson_2010GReGr_1,mathisson_2010GReGr_2} and further explored by \citet{Taub_1965}, \citet{tulczyjew1,tulczyjew1_1962}, and \citet{madore_1969}. However, the most significant advance in tackling this problem was achieved by Dixon\index{Dixon} \citep{dixon_1970_1,dixon_1970_2,dixon_1974_3,dixon_1973GReGr,dixon_1979} who followed Mathisson and worked out a more rigorous mathematical theory of covariant equations of motion of extended bodies starting from the microscopic law of conservation of matter,  
\be\la{wk10}
\nabla_\a T^{\a\b}=0\;,
\ee
where $\nabla_\a$ denotes a covariant derivative on spacetime manifold $M$ with metric $g_{\a\b}$, and $T^{a\b}$ is the stress-energy tensor\index{tensor!stress-energy} of matter of the extended bodies. Mathisson has dubbed this approach to the derivation of covariant equations of motion as {\it variational dynamics} \citep{mathisson_2010GReGr_1}. Dixon has advanced the original Mathisson's theory of variational dynamics. The generic approach used by Dixon\index{Dixon} was based on the introduction of the Riemann normal coordinates \citep[\S 11.6]{mtw} and the formalism of two-point world function\index{world function} $\sigma(z,x)$ and its partial derivatives (called sometimes bi-tensors\index{bi-tensor}) introduced by Synge \citep{syngebook}, the distributional theory of multipoles stemmed from the theory of generalized functions \citep{Gelfand_1964}, and the fiber bundle theory with horizontal and vertical (or Ehresmann's \citep{Kolar_1993}) covariant derivatives of two-point tensors. Dixon uses a vector bundle formed by the direct product of a reference time-like worldline ${\cal Z}$ considered as a base with a fiber being a space-like hypersurface consisting of geodesics emitted from point $z$ on ${\cal Z}$ in all directions orthogonal to ${\cal Z}$. Among other geometric structures extensively used in Dixon's approach to the variational dynamics, are Veblen's tensor extensions \citep{Veblen_1923} which allow to convert any geometric equation derived in the Riemann normal coordinates to the covariant form being valid in arbitrary coordinates. 

An extended body in Dixon's approach is idealized as a time-like world tube filled up with continuous matter which stress-energy tensor $T^{\a\b}$ vanishes outside the tube. By making use of the bi-tensor propagators, $K^\a{}_\m\equiv K^\a{}_\m(z,x) $ and $H^\a{}_\m\equiv H^\a{}_\m(z,x)$, composed out of the inverse matrices of the first-order partial derivatives of the world function $\sigma(z,x)$ with respect to $z$ and $x$, Dixon defined the total linear momentum\index{linear momentum}, $p^\a\equiv p^\a(z)$, and the total angular momentum\index{spin}, $S^{\a\b}\equiv S^{\a\b}(z)$, of the extended body by integrals over a space-like hypersurface $\Sigma$, \citep[Equations 66--67]{dixon_1979}
\ba\la{wk11}
{p}^\a&\equiv&\int\limits_\Sigma K^\a{}_\m T^{\m\n}\sqrt{- g}d\Sigma_\n\;,\\
\la{wk12}
S^{\a\b}&\equiv&-2\int\limits_\Sigma X^{[\a}H^{\b]}{}_\m T^{\m\n}\sqrt{- g}d\Sigma_\n\;,
\ea
where $z\equiv z^\a(\tau)$ is a reference worldline $\mathcal{Z}$ of a representative point that is assumed to be a center of mass\index{center of mass} of the body with $\tau$ being the proper time on this worldline, vector 
\be\label{wk12777}
X^\a=-{g}^{\a\b}(z)\frac{\pd\sigma(z,x)}{\pd z^\b}\;,
\ee
is tangent to a geodesic emitted from the point $z$ and passing through a field point $x$. The oriented element of integration on the hypersurface, 
\ba\label{pd6s}
d\Sigma_\a=\frac1{3!}E_{\a\m\n\s} dX^\m\wedge dX^\n\wedge dX^\s\;,
\ea
where $E_{\a\m\n\s}$ is 4-dimensional, fully anti-symmetric symbol of Levi-Civita, and the symbol $\wedge$ denotes the wedge product \citep[\S 3.5]{mtw} of the 1-forms $dX^\a$. Notice that Dixon's definition \eqref{wk12} of $S^{\a\b}$ has an opposite sign as compared to our definition \eqref{spin-9} of spin.

It is further assumed in Dixon's formalism that the linear momentum, ${p}^\a$, is proportional to the {\it dynamic} velocity, ${\mathfrak n}^\a$, of the body \citep[Equation 83]{dixon_1979}
\be\label{q13m} 
{p}^\a\equiv m{\mathfrak n}^\a\;,
\ee
where $m=m(\tau)$ is the total mass of the body which, in general, can depend on time. The {\it dynamic} velocity is a unit vector, ${\mathfrak n}_\a {\mathfrak n}^\a=-1$. The {\it kinematic} 4-velocity of the body moving along worldline $\mathcal{Z}$ is tangent to this worldline, $ u^\a=dz^\a/d\tau$. It relates to the {\it dynamic} 4-velocity by condition, ${\mathfrak n}_\a { u}^\a=-1$, while the normalization condition of the {\it kinematic} 4-velocity is ${ u}_\a { u}^\a=-1$. Notice that in the most general case the dynamic and kinematic velocities are not equal due to the gravitational interaction between the bodies of $N$-body system -- see \citep[Equation 88]{dixon_1979} and \citep{ehlers_1977GReGr} for more detail. 

Dixon defines the mass dipole, $m^\a=m^\a(z,\Sigma)$, of the body \citep[Equations 78]{dixon_1979},
\be 
\label{q14m}
m^\a\equiv S^{\a\b}{\mathfrak n}_\b\;,
\ee
and chooses the worldline $z=z^\a(\tau)$ of the center of mass of the body by condition, $m^\a=0$. This condition is equivalent due to (\ref{q13m}) and \eqref{q14m}, to  
\be\la{q13am}
{p}_\b S^{\a\b}=0\;,
\ee
which is known as Dixon's supplementary condition \index{Dixon's supplementary condition}\citep[Equation 81]{dixon_1979}. 

Dixon builds the body-adapted, local coordinates at each point $z$ on worldline ${\cal Z}$ as a set of the Riemann normal coordinates \citep[Chapter III, \S 7]{Schouten_book} denoted by $X^\a$ with the time coordinate $X^0$ along a time-like geodesic in the direction of the {\it dynamic} velocity ${\mathfrak n}^\a$, and the spatial coordinates $X^i=\{X^1,X^2,X^3\}$ lying on the hypersurface $\Sigma=\Sigma(z)$ consisting of all space-like geodesics passing through $z$ orthogonal to the unit vector ${\mathfrak n}^\a$ so that,
\be\label{be4y6}
{\mathfrak n}_\a X^\a=0\;.
\ee
It is important to understand that the Fermi normal coordinates (FNC) of observer moving along time-like geodesic do not coincide with the Riemann normal coordinates (RNC) used by Dixon \citep{dixon_1979,dixon_2008}. The FNC are constructed under condition that the Christoffel symbols vanish at {\it every} point along the geodesic \citep[Chapter III, \S 8]{Schouten_book} while the Christoffel symbols of the RNC vanish only at  a single event on spacetime manifold $M$. The correspondence between the RNC and the FNC is discussed, for example, in \citep[Chapter 5]{poissonwill_book}, \citep{nesterov_1999CQG} and generalization of the FNC for the case of accelerated and locally-rotating observers is given in \citep[\S 13.6]{mtw} and \citep{Ni_1978PRD}. The present paper uses the harmonic gauge, defined below in \eqref{hhb3vx5z}, to build the body-adapted local coordinates which coincide with the FNC of accelerated observer in the linearized approximation of the Taylor expansion of the metric tensor done with respect to the spatial coordinates around the worldline of the observer. 

Further development of the variational dynamics requires a clear separation of the matter and field variables in the solution of the full Einstein's field equations. This problem has not been solved in the MPD approach explicitly. It was replaced with the solution of a simpler problem of the separation of the matter and field variables in the equations of motion \eqref{wk10} by introducing a symmetric tensor distribution $\hat T^{\m\n}$ known as the stress-energy {\it skeleton} of the body \citep{mathisson_2010GReGr_1,mathisson_2010GReGr_2,dixon_1979}. Effectively, it means that the variational dynamics of each body is described on the effective background manifold $\bar M$ that is constructed from the full manifold $M$ by removing from the metric the self-field effects of body B. We denote the geometric quantities and fields defined on the effective background manifold $\bar M$ with a bar above the corresponding object. Mathematical construction of the effective background manifold $\bar M$ in our formalism is given below in section \ref{po3v6}.

\citet[Equation 140]{dixon_1979} defined high-order multipoles of an extended body in the normal Riemann coordinates, $X^\a$, by means of a tensor integral   
\be\label{q12m} 
I^{\a_1...\a_l\m\n}(z)=\int X^{\a_1}...X^{\a_l}\hat T^{\m\n}(z,X)\sqrt{-\bar g(z)}DX\;,\qquad\qquad (l\ge 2)
\ee
where the coordinates $X^\a$ are connected to the Synge world function $\sigma$ in accordance to (\ref{wk12777}), $\hat T^{\m\n}$ is the stress-energy {\it skeleton} of the body, and the integration is performed over the tangent space of the point $z$ with the volume element of integration $DX=dX^0\wedge dX^1\wedge dX^2\wedge dX^3$. Dixon's multipoles have specific algebraic symmetries which significantly reduce the number of linearly-independent components of $I^{\a_1...\a_l\m\n}$. These symmetries are discussed in section \ref{appndxon} of the present paper. 

\citet{dixon_1979} presented a number of theoretical arguments suggesting that the covariant equations of motion of the extended body have the following covariant form \citep[Equations 4.9--4.10]{dixon_1973GReGr}
\ba\la{q15m}
\frac{{\cal D}{p}_\a}{{\cal D}\tau}&=&\frac12 {u}^\b S^{\m\n}\bar R_{\m\n\b\a}+\frac12\sum\limits_{l=2}^{\infty}\frac1{l!}\nabla_{\a} A_{\b_1...\b_l\m\n}I^{\b_1...\b_l\m\n}
\\\la{q16m}
\frac{{\cal D}S^{\a\b}}{{\cal D}s}&=&2{p}^{[\a}{u}^{\b]}+\sum\limits_{l=1}^{\infty}\frac{1}{l!}{B}_{\g_1...\g_l\s\m\n}\bar g^{\s[\a}I^{\b]\g_1...\g_l\m\n}\;,
\ea
where ${\cal D}/{\cal D}\tau\equiv {u}^\a\nabla_\a$
is the covariant derivative taken along the reference line $z=z(\tau),$ the moments $I^{\a_1...\a_l\m\n}$ are defined in \eqref{q12m}, $A_{\b_1...\b_l\m\n}$ and ${B}_{\g_1...\g_l\s\m\n}$ are the symmetric tensors computed at point $z$, and the bar above any tensor indicates that it belongs to the effective background manifold $\bar M$. 

\citet{th_1985} call body's multipoles $I^{\a_1...\a_l\m\n}$ the {\it internal} multipoles. Tensors $A_{\b_1...\b_l\m\n}$ and $B_{\g_1...\g_l\m\n\s}$ are called the {\it external} multipoles of the background spacetime. The external multipoles are the {\it normal} tensors in the sense of \citet{Veblen_1923}. They are reduced to the repeated partial derivatives of the metric tensor, $\bar g_{\m\n}$, and the Christoffel symbols, $\bar\Gamma_{\s\m\n}$, in the Riemann normal coordinates taken at the origin of the coordinate $X=0$ (corresponding to the point $z$ in coordinates $x^\a$) \citep{dixon_1979,Schouten_book},
\ba\label{om5g}
A_{\b_1...\b_l\m\n}&=&\lim_{X\rightarrow 0}\pd_{\b_1...\b_l}\bar g_{\m\n}(X)\;,\\
\label{om6f}
B_{\b_1...\b_l\s\m\n}&=&2\lim_{X\rightarrow 0}\pd_{\b_1...\b_l}\Gamma_{\s\m\n}(X)\\\nonumber
&=&\lim_{X\rightarrow 0}\left[\pd_{\b_1...\b_l\s}\bar g_{\m\n}(X)+\pd_{\b_1...\b_l\m}\bar g_{\n\s}(X)-\pd_{\b_1...\b_l\n}\bar g_{\s\m}(X)\right]\;.
\ea
In arbitrary coordinates $x^\a$, the normal tensors are expressed in terms of the Riemann tensor, $\bar R^\a{}_{\m\b\n}$, and its covariant derivatives \citep[Chapter III, \S 7]{Schouten_book}. More specifically, if the terms being quadratic with respect to the Riemann tensor are neglected, the external Dixon multipoles read,
\ba\label{zxc3d}
A_{\b_1...\b_l\m\n}&=&2\frac{l-1}{l+1}\nabla_{(\b_1...\b_{l-2}}\bar R_{|\m|\b_{l-1}\b_l)\n}\;,\\\label{ydrv3}
B_{\b_1...\b_l\s\m\n}&=&\frac{2l}{l+2}\left[\nabla_{(\b_1...\b_{l-1}}\bar R_{|\m|\s\b_l)\n}+\nabla_{(\b_1...\b_{l-1}}\bar R_{|\s|\m\b_l)\n}-\nabla_{(\b_1...\b_{l-1}}\bar R_{|\s|\n\b_l)\m}\right]
\ea  
where the vertical bars around an index means that it is excluded from the symmetrization denoted by the round parentheses. Notice that each term with the Riemann tensor in \eqref{zxc3d}, \eqref{ydrv3} is symmetric with respect to the first and fourth indices of the Riemann tensor. This tells us that $A_{\b_1...\b_l\m\n}=A_{(\b_1...\b_l)(\m\n)}$ and $B_{\g_1...\g_l\s\m\n}=B_{(\g_1...\g_l)(\s\m)\n}$ in accordance with the symmetries of \eqref{om5g}, \eqref{om6f}.

Substituting these expressions to \eqref{q15m}, \eqref{q16m} yields the Dixon equations of motion in the following form, 
\ba\la{q15ms}
\frac{{\cal D} {p}_\a}{{\cal D}\tau}&=&\frac12 {u}^\b S^{\m\n}\bar R_{\m\n\b\a}+\sum\limits_{l=2}^{\infty}\frac{l-1}{(l+1)!} \nabla_{\a(\b_1...\b_{l-2}}\bar R_{|\m|\b_{l-1}\b_l)\n} J^{\b_1...\b_{l-1}\m\b_l\n}\;,
\\\la{q16mc}
\frac{{\cal D} S^{\a\b}}{{\cal D}\tau}&=&2{p}^{[\a}{u}^{\b]}+2\sum\limits_{l=1}^{\infty}\frac{l(l+1)}{(l+2)!}\nabla_{(\g_1...\g_{l-1}}\bar R_{|\m|\s\g_l)\n}\bar g^{\s[\a} J^{\b]\g_1...\g_{l-1}\m\g_l\n}\;,
\ea
where 
\be\la{wk12b}
J^{\a_1...\a_{p}\l\m\s\n}\equiv I^{\a_1...\a_{p}[\l[\s\m]\n]}\;,
\ee
denotes the internal multipoles with a skew symmetry with respect to two pairs of indices, $[\l\m]$ and $[\s\n]$. The interrelation between the Dixon  $I$ and $J$  multipoles is explained in more detail in section \ref{appndxon} of the present paper. 

Mathematical elegance and apparently covariant nature of the variational dynamics has been attracting researchers to work on improving various aspects of derivation of the MPD equations of motion  \citep{beig_1967CMaPh,schattner_1979GReGr,ehlers_1977GReGr,Ehlers_1980NYASA,bailey_1980AnPhy,steinhoff_2010PhRvD,dirk_obukhov2014,Obukhov_Puetzfeld2014,Pound_2015}. From astrophysical point of view Dixon's formalism is viewed as being of considerable importance for the modeling the gravitational waves emitted by the extreme mass-ratio inspirals (EMRIs) which are binary black holes consisting of a super-massive black hole and a stellar mass black hole. EMRIs form a key science goal for the planned space based gravitational wave observatory LISA and the equations of motion of the black holes in those systems must be known with unprecedented accuracy \citep{Schutz_2018RSPTA,Babak2015}. Nonetheless, in spite of the power of Dixon's mathematical apparatus, there are several issues which make the Dixon theory of the variational dynamics yet unsuitable for relativistic celestial mechanics, astrophysics and gravitational wave astronomy. 

The main problem is that the variational dynamics is too generic and does not engage any particular theory of gravity. It tacitly assumes that some valid theory of gravity is chosen, gravitational field equations are solved, and the metric tensor is known. However, the field equations and the equations of motion of matter are closely tied up -- matter generates gravity while gravity governs motion of matter. Due to this coupling the definition of the center of mass, linear momentum, spin, and other body's internal multipoles depend on the metric tensor which, in its own turn, depends on the multipoles through the non-linearity of the field equations. It complicates the problem of interpretation of the gravitational stress-energy skeleton in the non-linear regime of gravitational field and makes the MPD equations \eqref{q15m}, \eqref{q16m} valid solely in the linearized approximation of general relativity. For the same reason it is difficult to evaluate the residual terms in the existing derivations of the MPD equations and their multipolar extensions. One more serious difficulty relates to the lack of prescription for separation of self-gravity effects of moving body from the external gravitational environment. The MPD equations of motion are valid on the effective background manifold $\bar M$ but its exact mathematical formulation remains unclear in the framework of the variational dynamics alone \citep{Pound_2015}. Because of these shortcomings the MPD variational dynamics has not been commonly used in real astrophysical applications in spite that it is sometime claimed as a "standard theory" of the equations of motion of massive bodies in relativistic gravity \citep{bini_2009GReGr}. 

In order to complete the MPD approach to variational dynamics and make it applicable in astrophysics several critical ingredients have to be added. More specifically, what we need are:
\begin{enumerate}
\item the procedure of unambiguous characterization and determination of the gravitational self-force and self-torque exerted by the body on itself, and the proof that they are actually vanishing;
\item the procedure of building the effective background manifold $\bar M$ with the background metric $\bar g_{\a\b}$ used to describe the motion of the body which is a member of ${N}$-body system;
\item the precise algorithm for calculating the body's internal multipoles (\ref{q12m}) and their connection to gravitational field of the body;
\item the relationship between the Blanchet-Damour mass and spin multipoles, ${\cal M}^{\a_1...\a_l}$ and ${\cal S}^{\a_1...\a_l}$ and the Dixon multipoles (\ref{q12m}).
\item the procedure of selection of the center-of-mass worldline $\mathcal{Z}$ within each body.
\end{enumerate}

Present paper implement the formalism of relativistic reference frames in $N$-body system worked out by \citet{bk89} and \citet{dsx1} to derive covariant equations of motion of massive bodies with all Blanchet-Damour multipoles taken into account by making use of the mathematical technique proposed by \citet{th_1985}. It relies upon the construction of the effective background manifold $\bar M$ by solving the Einstein field equations and applying the asymptotic matching technique which separates the self-field effects from external gravitational environment, defines all external multipoles and establishes the local equations of motion of the body in the body-adapted local coordinates. The body's internal multipoles are defined in the harmonic gauge by solving the field equations in the body-adapted local coordinates as proposed by \citet{bld}. The covariant equations of motion follow immediately from the local equations of motion by applying the Einstein equivalence principle \citep{th_1985}. We compare our covariant equations of motion with the Dixon equations \eqref{q15ms}, \eqref{q16mc} in section \ref{appendixB}. 

\section{Atlas of Spacetime Manifold in N-body Problem}\label{qw231z}

We consider an isolated gravitating system consisting of $N$ extended bodies with continuous distribution of mass, velocity, and other functions characterizing their internal structure. The material variables are described by the stress-energy tensor $T^{\a\b}$ and the field variables are components of the metric tensor $g_{\a\b}$ obeying the Einstein field equations. Gravitational field of the whole $N$-body system can be described in a single coordinate chart $x^\a=(t,x^i)$ covering the entire spacetime manifold $M$ approaching asymptotically the Lorentzian coordinates of a flat spacetime. The global coordinates are indispensable for describing a relative motion of the bodies with respect to each other but they are notoriously unhelpful for solving the internal problem of motion of matter inside each body and for defining a set of multipoles characterizing its own gravitational field. It requires to introduce a local coordinate chart $w^\a=(u,w^i)$ adapted to each body. Hence, the entire manifold $M$ turns out to be covered by a set of $N$ local coordinates overlapping with each other and with the global coordinate chart. The set of $N+1$ coordinate charts form an atlas of the spacetime manifold $M$ which can be described in many different ways depending on the choice of the gauge conditions imposed on the solutions of the Einstein equations. One of the most convenient gauges is the harmonic gauge \citep{fockbook,weinberg_book1972}
\be\label{hhb3vx5z}
\pd_\b\left(\sqrt{-g} g^{\a\b}\right)=0\;,
\ee
which we use in the present paper. 

Notice that Dixon's equations of motion \eqref{q15ms}--\eqref{q16mc} have been derived in, what we call, the Riemann normal coordinate (RNC) gauge. The RNC gauge is defined by the condition that the Christoffel symbols and all its consecutive symmetric derivatives vanish at the origin in Riemann normal coordinates \citep[Chapter 3, \$ 7]{Schouten_book}, that is on the reference worldline ${\cal W}$ of the origin of the local coordinates adapted to body B
\be\label{rnc_gauge}
\pd_{(\a_1...\a_l)}\G^\b_{\m\n}\Big|_{\cal W}=0\;,
\ee
for any $l\ge 0$. 

Imposing a certain gauge condition eliminates the spurious, gauge-dependent components from the metric tensor which have no direct physical meaning. Vacuum gravitational field is fully characterized by only two components of the metric tensor which are characterized by two corresponding sets of multipoles. Choosing a gauge condition does not rigidly fix coordinates because each particular gauge condition admits a residual gauge freedom which corresponds to a different choice of coordinates within the chosen gauge. This is completely similar to the situation in electrodynamics or any other field theory where the choice of the gauge condition limits the numbers of physically-admissible solutions of the field equations but does not affect our freedom in choosing coordinates \citep{Petrov_2017book}. This remark is important for unambiguous understanding of physical meaning of covariant equations of motion of extended bodies which are valid in a specific gauge but in arbitrary coordinates.

The components of the metric tensor, $g_{\a\b}\equiv g_{\a\b}(t,{\bm x})$, in the global coordinates are given by equations \citep{fockbook,willbook}
\be\label{zra7c}
g_{\a\b}=\eta_{\a\b}+h_{\a\b}\;,
\ee
where
\ba\la{avzr23}
h_{00}&=&2U(t,{\bm x})-2U^2(t,{\bm x})-\pd_{tt}\chi(t,{\bm x})\;,\\
\la{azx8na}
h_{0i}&=&-4 U^i(t,{\bm x})\;,\\
\la{vdx8nb}
h_{ij}&=&2\delta_{ij}U(t,{\bm x})\;,
\ea
where $U$, $U^i$ and $\chi$ are scalar and vector potentials describing the gravitational field of {\it all} bodies of $N$-body system
\begin{equation}
  \label{12.9a}
  U(t,\bm{x})=\sum_{{\rm B}=1}^N U_{\rm B}(t,\bm{x}),\qquad U^i(t,\bm{x})=\sum_{{\rm B}=1}^N U^i_{{\rm B}}(t,\bm{x}),\qquad\chi(t,\bm{x})=\sum_{{\rm B}=1}^N\chi_{\rm B}(t,\bm{x})\;.
\end{equation}
Here, the gravitational potentials of body C are defined as integrals performed over a spatial volume ${\cal V}_{\rm B}$ occupied by matter of body B,
\begin{eqnarray}
\label{12.11}
  U_{\rm B}(t,\bm{x}) &=&\int\limits_{{\cal V}_{\rm B}}\frac{\sigma(t,\bm{x}')}{|\bm{x}-\bm{x}'|}d^3x'\;,
\\
\label{12.11eee}
  U^i_{\rm B}(t,\bm{x}) &=&\int\limits_{{\cal V}_{\rm B}}\frac{\sigma^i(t,\bm{x}')}{|\bm{x}-\bm{x}'|}d^3x'\;,
\\
  \label{12.12}
  \chi_{\rm B}(t,\bm{x}) &=& -\int\limits_{{\cal V}_{\rm B}}\sigma(t,\bm{x}')|\bm{x}-\bm{x}'|d^3x'\;,
\ea
where
\begin{eqnarray}
  \label{pz3q1}
  \sigma(t,\bm{x}) & = & \rho^{\ast}(t,\bm{x})\left[1+\frac32 v^2(t,\bm{x})+\Pi(t,\bm{x})
  - {U}_{\rm B}(t,\bm{x})  \right]+{\mathfrak{s}}^{kk}(t,\bm{x})\;,
\\  
\label{pz629}
  \sigma^i(t,\bm{x})&=&\rho^{\ast}(t,\bm{x})v^i(t,\bm{x}),
\ea
are mass and current densities of matter of body C referred to the global coordinates, $\rho^{\ast}=\rho\sqrt{-g}u^0$ is the invariant density of matter \citep{willbook}, $\rho$ is the local density of matter, $\Pi$ is the density of internal energy, ${\mathfrak{s}}^{ij}$ is the spatial stress energy tensor, and $v^i=dx^/dt$ is velocity of matter. It is useful to emphasize that all volume integrals defining the metric tensor in the global coordinates, are taken on the space-like hypersurface of constant coordinate time $t$.    

To a large extent each body B falls freely in the external gravitational field of the other $N-1$ bodies. Therefore, the metric tensor, $\rmg_{\a\b}\equiv\rmg_{\a\b}(u,{\bm w})$, in the local coordinates adapted to the body is a linear superposition of the solution of inhomogeneous Einstein equations with the stress-energy tensor of the body B and a general solution of the homogeneous Einstein equations describing the tidal field of the external bodies.
The metric in the local coordinates adapted to body B reads \citep{th_1985,Zhang_1985PhRvD}
\be\label{bvxcwe3}
\rmg_{\a\b}=\eta_{\a\b}+\rmh^{\rm B}_{\a\b}+\rmh^{\rm E}_{\a\b}+\rmh^{\rm I}_{\a\b}\;,
\ee
where
\ba\la{vzr23}
\rmh^{\rm B}_{00}&=&2U_{\rm B}(u,{\bm w})-2U_{\rm B}^2(u,{\bm w})-\pd_{uu}\chi_{\rm B}(u,{\bm w})\;,\\
\la{zx8na}
\rmh^{\rm B}_{0i}&=&-4 U_{\rm B}^i(u,{\bm w})\;,\\
\la{dx8nb}
\rmh^{\rm B}_{ij}&=&2\delta_{ij}U_{\rm B}(u,{\bm w})\;,
\ea
are the metric tensor perturbations describing gravitational field of body B, 
\begin{eqnarray}
  \label{1.24b}
  \rmh^{\rm E}_{00}(u,\bm{w}) & = &2\sum_{l=1}^{\infty}\frac{1}{l!}{\cal Q}_Lw^{L}-2\bigg(\sum_{l=1}^{\infty}\frac{1}{l!}{\cal Q}_Lw^{L}\bigg)^2  +\sum_{l=1}^{\infty}\frac{1}{(2l+3)l!}\ddot{\cal Q}_Lw^{L}w^2\;,\\
  \label{1.25ba}
  \rmh^{\rm E}_{0i}(u,\bm{w}) & = & 
\sum_{l=1}^{\infty}\frac{l+1}{(l+2)!}\varepsilon_{ipq}{\cal C}_{pL}w^{<qL>}
  +4\sum_{l=1}^{\infty}\frac{2l+1}{(2l+3)(l+1)!}\dot{\cal Q}_Lw^{<iL>}\;,
\\
  \label{1.26b}
  \rmh^{\rm E}_{ij}(u,\bm{w}) & = & 2\delta_{ij}\sum_{l=1}^{\infty}\frac{1}{l!}{\cal Q}_Lw^{L},
\end{eqnarray}
are the metric tensor perturbations describing the tidal gravitational field of the external $N-1$ bodies in the vicinity of the worldline of the origin of the local coordinates adapted to body B, and, here and anywhere else, the angular brackets around a group of tensor indices denote STF projection \eqref{stfformula}.
Finally,
\begin{eqnarray}
  \label{1.24o}
  \rmh^{\rm I}_{00} =  -4U_{\rm B}(u,{\bm w})\sum_{l=1}^{\infty}\frac{1}{l!}{\cal Q}_Lw^{L}
  -2\sum_{l=1}^{\infty}\frac{1}{l!}{\cal Q}_L\int\limits_{{\cal V}_{\rm B}}\frac{\rho^{\ast}(u,\bm{w}')w'^L}{|\bm{w}-\bm{w}'|}d^3w'\;,
\end{eqnarray}
is the metric tensor perturbation cause by the non-linear interaction of $\rmh^{\rm B}_{\a\b}$ and $\rmh^{\rm E}_{\a\b}$ through the Einstein equations. 

Gravitational potentials $U_{\rm B}(u,{\bm w})$, $U^i_{\rm B}(u,{\bm w})$, $\chi_{\rm B}(u,{\bm w})$ are given by equations \eqref{12.11}--\eqref{pz629} after replacement of the global coordinates to the local ones, and taking into account that integration in the local coordinates is performed on a hypersurface of constant coordinate time $u$. The gravitoelectric, ${\cal Q}_L$, and gravitomagnetic, ${\cal C}_L$, external multipoles are STF tensors with respect to all indices that is, ${\cal Q}_L\equiv {\cal Q}_{<i_1...i_l>}$ and ${\cal C}_L\equiv {\cal C}_{<i_1...i_l>}$.

\section{Matching the Global and Local Charts}\label{nrv27x}  

Global and local coordinates are interconnected through the tensor law of transformation of the metric tensor perturbations, 
\begin{eqnarray}
  \label{2.6}
  h_{\mu\nu}(t,\bm{x}) & = & \rmh_{\alpha\beta}(u, \bm{w})\frac{\partial w^{\alpha}}{\partial x^{\mu}} \frac{\partial w^{\beta}}{\partial x^{\nu}}\;.
 \end{eqnarray}
Equation(\ref{2.6}) matches the gravitational field variables in the spacetime region covered by both the local and global coordinates. Metric perturbations in the left-hand side of this equations are given by integrals performed over volumes of all bodies of $N$-body system on hypersurface of constant time $t$. The right-hand side of \eqref{2.6} contains, besides the integrals from the matter variables of body B taken on hypersurface of constant time $u$, the external multipoles ${\cal Q}_L$, ${\cal C}_L$ from the external part of the metric tensor in the local coordinates and yet unknown transformation functions $w^\a=w^\a(x^\b)$. Substituting the metric perturbations \eqref{zra7c} and \eqref{bvxcwe3} to the left and to the right hand sides of \eqref{2.6} respectively we find out that all terms which depend on the internal potentials of body B (and which multipolar expansions are singular at the origin of the local chart) are canceled out identically in the matching equation \eqref{2.6}.

Solving \eqref{2.6} for the remaining terms allows to determine the multipoles and the transformation functions along with equations of motion of the origin of the local coordinates. The solution is given by the following equations \citep{Kopejkin_1988CeMec,kopejkin_1991INTSA,bk-nc},
\begin{eqnarray}
  \label{5.12}
  u & = & t+\mathcal{A}-v^k_{\rm B}R_{\rm B}^k+\left(\frac{1}{3}v^k_{\rm B}a^k_{\rm B}-\frac{1}{6}\dot{\bar{U}}(t,{\bm x}_\B)- \frac{1}{10}\dot{a}^k_{\rm B}R^k_{\rm B}\right)R^2_{\rm B}+\sum_{l=1}^{\infty}\frac{1}{l!}\mathcal{B}^LR_{\rm B}^L\;,\\
  \label{5.13}
  w^i & = & R^i_{\rm B}+\left(\frac{1}{2}v^i_{\rm B}v_{\rm B}^k+\delta^{ik}\bar{U}(t,{\bm x}_\B)+F^{ik}_\B\right)R_{\rm B}^k+ a^k_{\rm B}R^i_{\rm B}R^k_{\rm B}-\frac{1}{2}a^i_{\rm B}R_{\rm B}^2\;,
\end{eqnarray}
where $R^i_{\rm B}=x^i-x^i_{\rm B}$
is the coordinate distance on the hypersurface of constant time $t$ between the field point, $x^i$, the origin of the local coordinates $x^i_{\rm B}\equiv x^i_{\rm B}(t)$, its velocity $v^i_{\rm B}\equiv dx^i_{\rm B}/dt$, and acceleration $a^i_{\rm B}\equiv dv^i_{\rm B}/dt$.
 
Function $\mathcal{A}$ defines transformation between the local coordinate time $u$ and the global coordinate time $t$ at the origin of the local coordinates. It obeys the ordinary differential equation \citep{Kopejkin_1988CeMec,bk_1990CeMDA},
\begin{eqnarray}
  \label{5.14}
  \frac{d\mathcal{A}}{dt} & = & -\frac{1}{2}v_{\rm B}^2
   - \frac{1}{8}v^4_{\rm B}-\bar{U}(t,{\bm x}_\B)-\frac{3}{2}v_{\rm B}^2\bar{U}(t,{\bm x}_\B)+\frac{1}{2}\bar{U}^2(t,{\bm x}_\B)+4v_{\rm B}^k\bar{U}^k(t,{\bm x}_\B) +\frac{1}{2}\pd_{tt}\bar{\chi}(t,{\bm x}_\B)\;.
\end{eqnarray}
The other functions entering (\ref{5.12}), (\ref{5.13}) are defined by algebraic relations \citep{kovl_2004,kopeikin_2011book} 
\begin{eqnarray}
  \label{5.15}
  \mathcal{B}^i & = & 4\bar{U}^i(t,{\bm x}_\B)-3v^i_{\rm B}\bar{U}(t,{\bm x}_\B)-\frac{1}{2}v^i_{\rm B}v^2_{\rm B}\;,\\
  \label{5.16}
  \mathcal{B}^{ij} & = & 4\pd^{<i}\bar{U}^{j>}(t,{\bm x}_\B)-4v_{\rm B}^{<i}\pd^{j>}\bar{U}(t,{\bm x}_\B)+2a_{\rm B}^{<i}a_{\rm B}^{j>}\;,\\
  \label{5.17}
  \mathcal{B}^{iL} & = & 4\pd^{<L}\bar{U}^{i>}(t,{\bm x}_\B)-4v_{\rm B}^{<i}\pd^{L>}\bar{U}(t,{\bm x}_\B)\;, \qquad\qquad\qquad\qquad (l\ge 2),
\end{eqnarray}
where the angular brackets denote STF projection of indices, and the external (with respect to body B) potentials $\bar{U}$, $\bar{U}^i$, $\bar\chi$ are defined by
\be
\label{12.9aq2}
 \bar U(t,\bm{x})=\sum_{{\rm C}=1\atop {\rm C}\not={\rm B}}^N U_{\rm C}(t,\bm{x}),\qquad \bar U^i(t,\bm{x})=\sum_{{\rm C}=1\atop {\rm C}\not={\rm B}}^N U^i_{{\rm C}}(t,\bm{x}),\qquad\bar \chi(t,\bm{x})=\sum_{{\rm C}=1\atop {\rm C}\not={\rm B}}^N\chi_{\rm C}(t,\bm{x})\;,
\end{equation}
where the summation runs over all bodies of the system except of body B.
Notations $\bar{U}(t,{\bm x}_\B)$, $\bar{U}^i(t,{\bm x}_\B)$, and $\bar\chi(t,{\bm x}_\B)$ mean that the potentials are taken at the origin of the local coordinates adapted to body B at instant of time $t$.

The skew-symmetric matrix ${F}^{ij}_\B$ of rotation of spatial axes of the local coordinates with respect to the global ones, is a solution of the ordinary differential equation \citep{Kopejkin_1988CeMec,kovl_2004}
\begin{equation}
  \label{5.18}
  \frac{d{F}^{ij}_\B}{dt}=4\pd^{[i}\bar{U}^{j]}(t,{\bm x}_\B)+3v_{\rm B}^{[i}\pd^{j]}\bar{U}(t,{\bm x}_\B)+v_{\rm B}^{[i}{\cal Q}^{j]}\;.
\end{equation}
The first term in the right-hand side of (\ref{5.18}) describes the Lense-Thirring (gravitomagnetic) precession\index{precession!Lense-Thirring}\index{precession!gravitomagnetic} which is also called the dragging of inertial frames \citep{ciufolini_book}. The second term in the right-hand side of \eqref{5.18} describes the de-Sitter (geodetic) precession\index{precession!de-Sitter}\index{precession!geodetic}, and the third term describes the Thomas precession \index{precession!Thomas} depending on the local (non-geodesic) acceleration ${\cal Q}^i=\d^{ij}{\cal Q}_j$ of the origin of the local coordinates. The Lense-Thirring and geodetic precessions have been recently measured in GP-B gyroscope experiment \citep{GPB_2015CQGra} and by satellite laser ranging (SLR) technique \citep{ciufolini_2012NewA,Ciufolini_2016EPJC}. 

Besides the explicit form of the coordinate transformation \eqref{5.12}--\eqref{5.18}, the matching equation \eqref{2.6} yields equations of the worldline of the origin of the local coordinates adapted to body B \citep{kopejkin_1991INTSA,kovl_2004,kopeikin_2011book},
\begin{eqnarray}
  \label{5.8}
 a^i_{\rm B}& = & \pd^i\bar{U}(t,{\bm x}_\B)-{\cal Q}^i+F^{ij}_{\rm B}{\cal Q}_j
-\frac{1}{2}\pd_{tt}\pd^{i}{\bar\chi}(t,{\bm x}_\B)+4\dot{\bar{U}}^i(t,{\bm x}_\B)
-4v^j_{\rm B}\pd^i\bar{U}^{j}(t,{\bm x}_\B)
 -3v^i_{\rm B}\dot{\bar{U}}(t,{\bm x}_\B)\\\nonumber
 &-&4\bar{U}(t,{\bm x}_\B)\pd^i\bar{U}(t,{\bm x}_\B)+ v^2_{\rm B}\pd^i\bar{U}(t,{\bm x}_\B)
- v^i_{\rm B}v^j_{\rm B} \pd^j\bar{U}(t,{\bm x}_\B)  +\frac12 v^i_{\rm B}v^j_{\rm B}{\cal Q}_j+v^2_{\rm B}{\cal Q}^i+3{\cal Q}^i\bar U(t,{\bm x}_\B)\;,    \nonumber
\end{eqnarray}
where dot above function denotes a total derivative with respect to time $t$, ${\cal Q}^i=\d^{ij}{\cal Q}_j$ is a dipole term ($l=1$) in the external solution $\rmh^{E}_{00}$ (\ref{1.24b}) which is a local acceleration of the worldline ${\cal W}$ with respect to geodesic. Notice that so far \eqref{5.8} does not yield equations of motion of the center of mass of body B. Its determination requires integration of the microscopic equations of motion of matter of body B in the body-adapted local coordinates.

\section{The Internal and External Multipoles of Each Body in N-body System}\label{km2va}

There are two families of the {\it canonical} internal multipoles in general relativity which are called mass and spin multipoles \citep{bld,bld1986,iau2000}.  The internal STF mass multipoles of body B, ${\cal M}^L\equiv {\cal M}^{<i_1i_2\ldots i_l>}$ for $l\ge 0$, are defined by equation \citep{kovl_2004,kopeikin_2011book}
\begin{eqnarray}\label{1.31}
  {\cal M}^{L} & = & \int\limits_{{\cal V}_{\rm B}}\sigma(u,\bm{w})\left(
  1-\sum_{k=1}^{\infty}\frac{1}{k!}{\cal Q}_{K}w^{<K>}\right)w^{<L>}d^3w
  +\frac{1}{(2l+3)}\left(\frac12\ddot{\cal N}^{<L>}-4\frac{2l+1}{l+1} \dot{\cal R}^{<L>} \right)
\end{eqnarray}
where the angular brackets around spatial indices denote STF Cartesian tensor \citep{thor,bld1986}, and 
\ba
  \label{NL15}
  \mathcal{N}^{L} &=& \int\limits_{{\cal V}_{\rm B}}\sigma(u,\bm{w})w^2w^{<L>}d^3w\;,\\
  \label{spin-85}
  {\cal R}^{L}&=&\int\limits_{{\cal V}_{\rm B}}\sigma^i(u,\bm{w}) w^{<iL>}d^3w\;,
\ea
are two additional {\it non-canonical}  sets of STF multipoles, ${\cal V}_{\rm B}$ is volume of body B over which the integration is performed. 
The mass density $\sigma$ in \eqref{1.31} is \citep{kovl_2004} (do not confuse it with the Synge world function $\sigma$ mentioned earlier)
\begin{eqnarray}
  \label{pz3}
  \sigma(u,\bm{w}) & = & \rho^{\ast}(u,\bm{w})\left[1+\frac32\nu^2(u,\bm{w})+\Pi(u,\bm{w})
  - {U}_{\rm B}(u,\bm{w})  \right]+{\mathfrak{s}}^{kk}(u,\bm{w})\;,
\end{eqnarray}
and vector function
\be
  \label{pz6}
  \sigma^i(u,\bm{w})=\rho^{\ast}(u,\bm{w})\nu^i(u,\bm{w}),
\ee
is matter's current density of body B referred to the local coordinates, $\rho^{\ast}=\rho\sqrt{-g}u^0$ is the invariant density of matter, $\rho$ is the local density of matter, $\Pi$ is the density of internal energy, ${\mathfrak{s}}^{ij}$ is the spatial stress energy tensor, and $\nu^i=dw^i/du$ is velocity of matter in the local coordinates \footnote{In the linear approximation $\nu^i(u,{\bm w})=v^i(t,{\bm x})-v^i_{\rm B}(t)+{\cal O}(c^{-2})$.}. All volume integrals defining the metric tensor in the local coordinates, are taken on the space-like hypersurface of constant coordinate time $u$.   

The internal STF spin multipoles of body B, ${\cal S}^L\equiv {\cal S}^{<i_1i_2\ldots i_l>}$ for $l\ge 1$, are defined by expression \citep{bld,kovl_2004}
\begin{equation}
  \label{1.32}
  \mathcal{S}^{L} = \int\limits_{{\cal V}_{\rm B}}\varepsilon^{pq<i_l} w^{i_{l-1}...i_1>p}\sigma^q(u,{\bm w})d^3w,
\end{equation}
where matter's current density $\sigma^q$ has been defined in (\ref{pz6}). Integrals (\ref{1.31})--\eqref{spin-85}, \eqref{1.32} are performed over hypersurface of a constant coordinate time $u$ and, hence, all multipoles of body B are functions of time $u$ only. They are STF Cartesian tensors in the tangent Euclidean space attached to the worldline $\cal W$ of the origin of local coordinates adapted to body B. Definition \eqref{1.32} is sufficient for deriving the post-Newtonian translational equations of motion of the extended bodies in ${N}$-body system. However, derivation of the post-Newtonian rotational equations of motion requires a post-Newtonian definition of the body's angular momentum (spin). We shall discuss it later.

The external STF multipoles of body B also form two families - gravitoelectric multipoles ${\cal Q}_L$ and gravitomagnetic multipoles ${\cal C}_L$.
External gravitoelectric multipoles ${\cal Q}_L\equiv {\cal Q}_{<i_1i_2\ldots i_l>}$ for $l\ge2$ are obtained by solving \eqref{2.6} and given by the following equation \citep{kovl_2004,kopeikin_2011book} 
\begin{eqnarray}
  \label{5.9}
  {\cal Q}^L & = & \pd^{<L>}\bar{U}(t,{\bm x}_\B)
  -\frac{1}{2}\pd_{tt}\pd^{<L>}\bar{\chi}(t,{\bm x}_\B)+4\pd^{<L-1}\dot{\bar{U}}^{i_l>}(t,{\bm x}_\B)
  - 4v_{\rm B}^j\pd^{<L>}\bar{U}^j(t,{\bm x}_\B)\\
  & +&(l-4)v_{\rm B}^{<i_l}\pd^{L-1>}\dot{\bar{U}}(t,{\bm x}_\B) +2v_{\rm B}^2\pd^{<L>}\bar{U}(t,{\bm x}_\B) -\frac{l}{2}v^j_{\rm B}v_{\rm B}^{<i_l}\pd^{L-1>j}\bar{U}(t,{\bm x}_\B)-l\bar{U}(t,{\bm x}_\B)\pd^{<L>}\bar{U}(t,{\bm x}_\B)\nonumber\\
  & -&  (l^2-l+4)a_{\rm B}^{<i_l}\pd^{L-1>}\bar{U}(t,{\bm x}_\B)-lF^{j<i_l}_{\rm B}\pd^{L-1>}\bar{U}^j(t,{\bm x}_\B)+X^L\;\nonumber,\qquad\qquad(l\ge2)
\end{eqnarray}
where 
\begin{eqnarray}
  \label{1q4d}
  X^L \equiv \left\{ \begin{array}{cc}
         3a_{\rm B}^{<i_1}a_{\rm B}^{i_2>} &\qquad\qquad \mbox{if $l=2$};\\
         &\\
        0 & \qquad\qquad\mbox{if $l\geq3$}.\end{array} \right. 
\end{eqnarray}
External gravitomagnetic multipoles ${\cal C}_L\equiv {\cal C}_{<i_1i_2\ldots i_l>}$ for $l\ge2$ are also obtained by solving \eqref{2.6} and given by \footnote{Formula \eqref{3.29} corrects a typo in \citep[Equation 5.74]{kopeikin_2011book} for the external gravitomagnetic multipole ${\cal C}_L$.}
\begin{eqnarray}
  \label{3.29}
  \varepsilon_{ipk}{\cal C}_{pL} & = & 8\biggl[ v_{\rm B}^{[i}\pd^{k]<L>}\bar U(t,{\bm x}_\B) +\pd^{<L>[i}\bar{U}^{k]}(t,{\bm x}_\B)
   -\frac{l}{l+1}\delta^{<i_{l}[i}\pd^{k]L-1>}\dot{\bar{U}}(t,{\bm x}_\B)\biggr]\;,\qquad (l\ge1)
\end{eqnarray}
where the dot denotes the time derivative with respect to time $t$, the angular brackets denote STF symmetry with respect to multi-index $L=i_1,i_2,\ldots,i_l$, and the square brackets denote anti-symmetrization: $T^{[ij]}=(T^{ij}-T^{ji})/2$. The external multipoles ${\cal Q}_L$ and ${\cal C}_L$ are STF tensors which are corresponding analogues of the Dixon external multipoles $A_{\a_1...\a_l\m\n}$ and $B_{\a_1...\a_l\m\n}$ introduced in (\ref{om5g}) and (\ref{om6f}). 

\section{Post-Newtonian Equations of Motion in the Local Chart}\label{mop3vx4}
\subsection{Translational Equations for Linear Momentum}\label{jgrt3x}
Mass of body B is a monopole moment defined by \eqref{1.31} for $l=0$. After making some transformations of the integrand we can bring the monopole term in \eqref{1.31} to the following form \citep{kopeikin_2011book}
\begin{eqnarray}
  \label{confmass}
  {\cal M} & = &M-\sum_{l=1}^{\infty}\frac{l+1}{l!}{\cal Q}_L{\cal M}^{L}\;,
\end{eqnarray} 
where
\begin{equation}
  \label{grmass}
  M = \int\limits_{{\cal V}_{\rm B}}\rho^{\ast}\bigg(1+\frac{1}{2}\nu^2+\Pi-\frac{1}{2}\hat{U}_{\rm B}\bigg)d^3w
\end{equation}
is a post-Newtonian mass of body B considered as fully-isolated from the external world \citep{willbook}, ${\cal M}^{L}$ are mass multipoles of the body defined in (\ref{1.31}). The last term in the right-hand side of \eqref{confmass} can be interpreted in the spirit of Mach's principle claiming that the body's inertial mass originates from its gravitational interaction with an external universe. Mach's idea is not completely right because the inertial mass of the body originates primarily from the mass $M$ of the body's matter. Nonetheless, it has a partial support as we cannot completely ignore the gravitational interaction of a single body with its external gravitational environment in the definition of the inertial mass of the body. This effect is important to take into account in inspiralling compact binaries as they are tidally distorted and, hence, the part of the inertial mass of each star associated with the very last term in \eqref{confmass} rapidly changes as the size of the binary shrinks. Time variation of the mass ${\cal M}$ is \citep{kopeikin_2011book}
\be \label{x1674} 
\dot{\cal M}=-\sum\limits_{l=1}^\infty\frac1{(l-1)!}\left({\cal Q}_L\dot{\cal M}^{L}+\frac{l+1}{l}\dot{\cal Q}_L{\cal M}^{L}\right)\;.
\ee

We define the post-Newtonian center of mass of each body B by equation (\ref{1.31}) taken for multipolar index $l = 1$,
\ba\la{brtv56b}
  {\cal M}^i & = &\int\limits_{{\cal V}_{\rm B}}\varrho(u,\bm{w})\left(1-\sum_{l=1}^{\infty}\frac{1}{l!}{\cal Q}_Lw^{L}\right)w^i\,d^3w- \frac{2}{5}\left(3 \dot{\cal R}^i-\frac14\ddot{\cal N}^i \right)\;,
\ea
The last two terms in the right-hand side of \eqref{brtv56b} can be transformed to 
\be\la{jnw8j}
\frac25\left(3 \dot{\cal R}^i-\frac14\ddot{\cal N}^i\right) =\int\limits_{{\cal V}_{\rm B}}\left(\rho^*\nu^2+{\mathfrak{s}}_{kk}-\frac12\rho^*\hat U_{\rm B}\right)w^id^3w+\sum_{l=1}^\infty \frac{1}{(l-1)!}{\cal Q}_L{\cal M}^{iL}-\frac12\sum_{l=0}^\infty\frac{1}{(2l+3)l!} {\cal Q}_{iL}{\cal N}^L\;.
\ee
Replacing \eqref{jnw8j} to \eqref{brtv56b} brings the mass dipole to the following form, 
\begin{eqnarray}
  \label{confdipole}
  {\cal M}^i& = & \int\limits_{{\cal V}_{\rm B}} \rho^{\ast}(u,\bm{w}) \left(1+\frac{1}{2}\nu^2+\Pi-\frac{1}{2}\hat{U}_{\rm B}\right)w^id^3w
  -\sum_{l=1}^{\infty}\frac{l+1}{l!}{\cal Q}_L{\cal M}^{iL}
  - \frac{1}{2}\sum_{l=0}^{\infty}\frac{1}{(2l+3)l!}{\cal Q}_{iL}\mathcal{N}^L \;,
\end{eqnarray}
where the STF {\it non-canonical} multipole, $\mathcal{N}^L$, has been defined in \eqref{NL15}. 

The linear momentum ${\mathfrak{p}}^i$ of body B is defined as the first derivative of the dipole \eqref{confdipole} with respect to the local time $u$,
\be\la{b5z0e}
{\mathfrak{p}}^i\equiv\dot{\cal M}^i\;,
\ee
where the overdot denotes the time derivative with respect to $u$.
After taking the time derivative from the dipole \eqref{confdipole} we obtain \citep{kopeikin_2011book},
\begin{eqnarray}
  \label{confmoment}
  {\mathfrak{p}}^i
  & = & \int\limits_{{\cal V}_{\rm B}}\rho^{\ast}\nu^i\bigg(1+\frac{1}{2}\nu^2+\Pi-\frac{1}{2}\hat{U}_{\rm B}\bigg)d^3w + \int\limits_{{\cal V}_{\rm B}}\left({\mathfrak{s}}_{ik}\nu^k-\frac{1}{2}\rho^{\ast}{W}^i_{\rm B}\right)d^3w-\sum_{l=1}^{\infty}\frac{1}{l!}{\cal Q}_L\int\limits_{{\cal V}_{\rm B}}\rho^{\ast}\nu^iw^{L}d^3w\\
  &-&\frac{d}{du}\bigg[\sum_{l=1}^{\infty}\frac{l+1}{l!}{\cal Q}_L{\cal M}^{iL} +\frac{1}{2}\sum_{l=0}^{\infty}\frac{1}{(2l+3)l!}{\cal Q}_{iL}\mathcal{N}^{L}\bigg]
  +\sum_{l=1}^{\infty}\frac{1}{l!}\bigg[{\cal Q}_L\dot{{\cal M}}^{iL}+\frac{l}{2l+1}{\cal Q}_{iL-1}\dot{\mathcal{N}}^{L-1}\bigg]\nonumber\;,
\end{eqnarray}
where 
\begin{equation}
  \label{w3h7}
  {W}^i_{\rm B} = \int\limits_{{\cal V}_{\rm B}}\frac{\rho^{\ast}(u,\bm{w}')\nu'^k(w^k-w'^k)(w^i-w'^i)}{|\bm{w}-\bm{w}'|^3}d^3w'\;,
\end{equation}
is a new potential of gravitational field of body B - c.f. \citep[Equation 4.32]{willbook}. We remind now that the point $x^i_{\rm B}$ represents position of the origin of the local coordinates adapted to body B in the global coordinates taken at instant of time $t$. It moves along worldline ${\cal W}$ which we want to make identical to worldline ${\cal Z}$ of the center of mass of body B.
It can be achieved if we can retain the center of mass of body B at the origin of the local coordinates adapted to the body, that is to have for any instant of time, 
\be\la{n5vz1o}
{\cal M}^i=0\;,\qquad\qquad {\mathfrak{p}}^i=0\;.
\ee
These constraints can be satisfied if, and only if, the local equation of motion of the center of mass of the body can be reduced to identity 
\begin{equation}
  \label{a0n5}
  \dot{\mathfrak{p}}^i(u) = \ddot{\cal M}^i\equiv 0\;.
\end{equation}

Equation \eqref{a0n5} can be fulfilled after making an appropriate choice of the external dipole ${\cal Q}_i$ that characterizes the acceleration of the origin of the local coordinates of body B. More specifically, the identity \eqref{a0n5} demands \citep{dsx2,kovl_2004}
\ba\la{q6v4m}
{\cal M}{\cal Q}_i&=&-\sum_{l=2}^{\infty}\frac{1}{l!}{\cal Q}_{iL}{\cal M}^{L}-\sum_{l=1}^{\infty}\frac{l}{(l+1)!}{\cal C}_{iL}\mathcal{S}^{L}\\\nonumber
&&+\sum_{l=1}^{\infty}\frac{l^2+l+4}{(l+1)!}{\cal Q}_L\ddot{{\cal M}}^{iL}
   + \sum_{l=1}^{\infty}\frac{2l+1}{l+1}\frac{l^2+2l+5}{(l+1)!}\dot{\cal Q}_L\dot{{\cal M}}^{iL}
  + \sum_{l=1}^{\infty}\frac{2l+1}{2l+3}\frac{l^2+3l+6}{(l+1)!}\ddot{\cal Q}_L{\cal M}^{iL}\\
& &+ \sum_{l=1}^{\infty}\frac{1}{(l+1)!}\varepsilon_{ipq}\bigg({\cal C}_{pL}\dot{{\cal M}}^{qL}+\frac{l+1}{l+2}\dot{\cal C}_{pL}{\cal M}^{qL}\bigg)
- \sum_{l=0}^{\infty}\frac{4}{l!(l+2)}\varepsilon_{ipq}\bigg({\cal Q}_{pL}\dot{\mathcal{S}}^{qL}
 + \frac{l+1}{l+2}\dot{\cal Q}_{pL}\mathcal{S}^{qL}\bigg)
\;.\nonumber
\ea
Equation \eqref{q6v4m} depends only on the {\it canonical} internal and external multipole moments. All {\it non-canonical} external moments have been removed by fixing the residual gauge freedom \citep{kovl_2004}. The {\it non-canonical} internal moments ${\cal R}^L$ and ${\cal N}^L$ have been absorbed to the post-Newtonian definition \eqref{1.31} of the internal {\it canonical} moments ${\cal M}^L$.  
Effectively, equation \eqref{q6v4m} is a relativistic analogue of the second Newton law for body B in the local coordinate chart which origin moves along the worldline ${\cal Z}$ of the center of mass of body B. This form of the post-Newtonian equations has been derived by Damour, Soffel and Xu (DSX) \citep{dsx2} in general relativity. Generalization of these equations to the case of scalar-tensor theory of gravity parameterized with two PPN parameters -- $\b$ and $\g$ \citep{willbook} -- has been given by \citet{kovl_2004}. We convert  equation \eqref{q6v4m}  to a fully-covariant form in section \ref{n3cz52s} and compare them with Dixon's equations of translational motion in section \ref{ndj45xc}. 

\subsection{Rotational Equations for Spin}

Spin ${\cal S}^i$ of an extended body B is defined in the Newtonian approximation by \eqref{1.32} taken for index $l=1$. It is insufficient for derivation of the post-Newtonian equations of rotational motion and must be extended to include the post-Newtonian terms. The post-Newtonian definition of spin can be extracted from the multipolar expansion of the metric tensor component $\rmh^{\rm B}_{0i}(u,{\bm w})$ by taking into account the post-post-Newtonian terms \cite{dyr2}. We have also to include the post-Newtonian terms from $\rmh^{\rm I}_{0i}(u,{\bm w})$ originating from the non-linear interaction of body B with the other bodies of $N$-body system. Such post-Newtonian definition of spin has been found in our work \citep{kopeikin_2011book}. It reads
\ba\label{spin-9}
{\cal S}^i &\equiv& 
\int\limits_{{\cal V}_{\rm B}}\rho^*\varepsilon_{ijk}w^j\nu^k\left(1+\frac12\nu^2+\Pi+3 U_{\rm B}\right)d^3w+\int\limits_{{\cal V}_{\rm B}}\varepsilon_{ijk}w^j{\mathfrak{s}}^{kp}\nu^pd^3w-
\frac1{2}\int\limits_{{\cal V}_{\rm B}}\rho^*\varepsilon_{ijk}w^j\bigg[
W^k_{\rm B} +7U^k_{\rm B}\bigg]d^3w\\\nonumber
&+&3\sum_{l=1}^{\infty}\frac1{l!}{\cal Q}_L\int\limits_{{\cal V}_{\rm B}}\rho^*\varepsilon_{ijk}w^j\nu^kw^Ld^3w
-\sum_{l=1}^{\infty}\frac{l}{(l+1)!}{\cal C}_L{\cal M}^{iL}
+\sum_{l=0}^{\infty}\frac{1}{(2l+3)l!}{\cal C}_{iL}{\cal
N}^{L}
\\\nonumber&+&
\frac12\sum_{l=0}^{\infty}\frac{1}{(2l+5)l!}\varepsilon_{ijk}\left[{\cal Q}_{kL}\dot{\cal
N}^{jL}- \frac{l+10}{l+2}\dot
{\cal Q}_{kL}{\cal N}^{jL}-8\frac{2l+3}{l+2}{\cal Q}_{kL}{\cal R}^{jL}\right]
\;,
\ea
where the {\it non-canonical}  multipoles, ${\cal N}^L$ and ${\cal R}^L$ have been defined earlier in \eqref{NL15} and \eqref{spin-85} respectively, $\nu^i=dw^i/du$ is velocity of matter of body B in the local coordinates, the integration is over volume of body B, and vector potential $W^k_{\rm B}$
is defined in (\ref{w3h7}).

Differentiation of \eqref{spin-9} with respect to the local time $u$ yields the rotational equation for spin of the body,
\be\label{spin-10}
\frac{d{\cal S}^i}{du}= {\cal T}^i \;,
\ee
where the torque
\be\label{spin-5} 
{\cal T}^i =\varepsilon_{ijk}
\sum_{l=1}^{\infty}\frac1{l!}\left({\cal Q}_{kL}
{\cal M}^{jL}+\frac{l+1}{l+2}{\cal C}_{kL}{\cal S}^{jL}\right)\;
\ee
We again observe that equation \eqref{spin-10} depends only on the {\it canonical} internal and external multipole moments. {\it Non-canonical} external moments have been removed by fixing the residual gauge freedom while the {\it non-canonical} internal moments ${\cal R}^L$ and ${\cal N}^L$ have been absorbed to the post-Newtonian definition \eqref{spin-9} of the internal angular momentum ${\cal S}^i$ and the internal mass multipoles ${\cal M}^L$.
The post-Newtonian spin-evolution equation \eqref{spin-10} is given in the local coordinates adapted to body B. It has been derived by Damour, Soffel and Xu (DSX) in \citep{dsx3} and extended to the case of scalar-tensor theory of gravity parameterized with two PPN parameters, $\b$ and $\g$, by \citet{kovl_2004}. We transform equations \eqref{spin-10}, \eqref{spin-5}  to a fully-covariant form in section \ref{mev254z4} and compare them with Dixon's equations for spin in section \ref{iopn3e4}.

\section{The Effective Background Manifold}\la{po3v6}      
Equations of translational motion of linear momentum \eqref{a0n5} and those of rotational motion for spin \eqref{spin-10} of an extended body B in the local coordinates depend on an infinite set of configuration variables which are the internal mass and spin multipoles of the body, ${\cal M}^L$ and ${\cal S}^L$, and the external gravitoelectric and gravitomagnetic multipoles, ${\cal Q}_L$ and ${\cal C}_L$. Each multipole is pinned down to worldline ${\cal Z}$ of the center of mass of the body. The equations of motion in the local coordinates can be lifted up to the generic covariant form by making use of the Einstein equivalence principle applied to body B that is treated as a massive particle endowed with the internal multipoles ${\cal M}^L$ and ${\cal S}^L$, and moving along the worldline ${\cal Z}$ on the effective background manifold $\bar M$ which properties are characterized by the external multipoles ${\cal Q}_L$ and ${\cal C}_L$ that are functions of the curvature tensor of the effective background manifold $\bar M$ and its covariant derivatives. The covariant form of the equations is independent of a particular realization of harmonic coordinates but we hold on the gauge conditions \eqref{hhb3vx5z} to prevent the appearance of gauge-dependent, nonphysical multipoles of gravitational field in the covariant equations of motion. 

The power of our approach to the covariant equations of motion is that unlike \citep{th_1985,ashb2} the effective background manifold $\bar M$ for each body B is not postulated or introduced ad hoc. It is constructed by solving the field equations in the local and global coordinate charts and separating the field variables in the internal and external parts. The separation is fairly straightforward in the local chart. The internal part of the metric tensor, $\rmh^{\rm B}_{\a\b}$, is determined by matter of body B and is expanded in the multipolar series outside the body which are singular at the origin of the body-adapted local coordinate chart. The external part of the metric tensor $\rmh^{E}_{\a\b}$ is a solution of vacuum field equations and, hence, is regular at the origin of the local chart. There are also internal-external coupling component $\rmh^{I}_{\a\b}$ of the metric tensor perturbation but its multipolar series is also singular at the origin of the local chart of body B. 

Spacetime geometry of the effective background manifold $\bar M$ is defined exclusively by the regular part of the metric tensor, $\bar\rmg_{\a\b}=\eta_{\a\b}+\rmh^{E}_{\a\b}$ since all terms which multipolar expansions are singular at the origin of the local chart cancel out identically in the matching equation \eqref{2.6}. This establishes a one-to-one correspondence between the external metric perturbation $\rmh^{E}_{\a\b}$ in the local chart and its counterpart in the global chart which is uniquely defined by the external gravitational potentials $\bar U, \bar U^i, \bar\chi$ given in (\ref{12.9aq2}). In section \ref{o8b4m1} we demonstrate that translational motion of the center of mass of body B can be interpreted as a perturbed geodesic of a massive particle on the effective background manifold $\bar M$ with the metric $\bar g_{\a\b}$. The particle has mass ${\cal M}$ and internal multipoles ${\cal M}^L$ and ${\cal S}^L$. The perturbation of the geodesic is caused by a local inertial force $F^i={\cal M}{\cal Q}^i$ arising due to the interaction of the particle's multipoles with the external gravitoelectric and gravitomagnetic multipoles, ${\cal Q}_L$ and ${\cal C}_L$, which are fully expressed in terms of the Riemann tensor $\bar R_{\a\b\m\n}$ of the background manifold $\bar M$ and its covariant derivatives. Covariant equations of rotational motion of the body spin are equations of the Fermi-Walker transport with the external torques caused by the coupling of the internal and external multipoles of the body.

The effective background metric $\bar g_{\a\b}$ is given in the global coordinates by the following equations, 
(cf. \citep{th_1985}),
\ba\la{v5x8n}
\bar g_{00}(t,{\bm x})&=&-1+2\bar U(t,{\bm x})-2\bar U^2(t,{\bm x})-\pd_{tt}\bar\chi(t,{\bm x})\;,\\
\la{v5x8na}
\bar g_{0i}(t,{\bm x})&=&-4\bar U^i(t,{\bm x})\;,\\
\la{v5x8nb}
\bar g_{ij}(t,{\bm x})&=&\delta_{ij}+2\delta_{ij}\bar U(t,{\bm x})\;,
\ea
where the potentials in the right hand side of (\ref{v5x8n})--(\ref{v5x8nb}) are defined in \eqref{12.9aq2} as functions of the global coordinates $x^\a=(t,{\bm x})$. The background metric in arbitrary coordinates can be obtained from (\ref{v5x8n})--(\ref{v5x8nb}) by performing a corresponding coordinate transformation. It is worth emphasizing that the effective metric $\bar g_{\a\b}$ is constructed for each body of the $N$-body system separately and is a function of the external gravitational potentials which depend on which body is chosen. It means that we have a collection of $N$ effective background manifolds $\bar M$ -- one for each extended body. Another prominent point to draw attention of the reader is the fact that the effective metric of the extended body B depends on the gravitational field of the body itself through the non-linear interaction \citep{Fichte_1950}.    

The background metric, $\bar g_{\a\b}$, is a starting point of the covariant development of the equations of motion. It has the Christoffel symbols
\be\la{qq2wkw}
\bar\G^\a_{\m\n}=\frac12\bar g^{\a\b}\left(\pd_\n\bar g_{\b\m}+\pd_\m\bar g_{\b\n}-\pd_\b\bar g_{\m\n}\right)\;,
\ee
which can be directly calculated in the global coordinates, $x^\a$, by taking partial derivatives from the metric components (\ref{v5x8n})--(\ref{v5x8nb}). In what follows, we shall make use of a covariant derivative defined on the background manifold $\bar M$ with the help of the Christoffel symbols $\bar\G^\a_{\m\n}$. The covariant derivative on the effective background manifold $\bar M$ is denoted $\bar\nabla_\a$ in order to distinguish it from the covariant derivative $\nabla_\a$ defined on the original spacetime manifold $M$. For example, the covariant derivative of vector field $V^\a$ is defined on the background manifold $\bar M$ by the following equation
\be\la{r5v6h}
\bar\nabla_\b V^\a=\pd_\b V^\a+\bar\G^\a_{\m\b}V^\m\;,
\ee 
which is naturally extended to tensor fields of arbitrary type and rank in a standard way \citep{kopeikin_2011book}.
It is straightforward to define other geometric objects on the background manifold $\bar M$ like the Riemann tensor, 
\be\la{b6c2j}
\bar R^\a{}_{\m\b\n}=\pd_\b\bar\G^\a_{\m\n}-\pd_\n\bar\G^\a_{\m\b}+\bar\G^\a_{\s\b}\bar\G^\s_{\m\n}-\bar\G^\a_{\s\n}\bar\G^\s_{\m\b}\;,
\ee
and its contractions -- the Ricci tensor $\bar R_{\m\n}=\bar R^\a{}_{\m\a\n}$, and the Ricci scalar $\bar R=\bar g^{\m\n}\bar R_{\m\n}$. Tensor indices on the background manifold $\bar M$\index{manifold!background} are raised and lowered with the help of the metric $\bar g_{\a\b}$. 

The background metric tensor in the local coordinates adapted to body B is given by 
\be\la{k8b2d}
{\bar {\rm g}}_{\a\b}(u,{\bm w})=\eta_{\a\b}+ \rmh^{E}_{\a\b}(u,{\bm w})\;,
\ee
where the perturbation $\hat \rmh^{E}_{\a\b}$ is given by the polynomial expansions \eqref{1.24b}--\eqref{1.26b} of the external gravitational field with respect to the local spatial coordinates.
Notice that at the origin of the local coordinates, where $w^i=0$, the background metric $\bar g_{\a\b}$ is reduced to the Minkowski metric $\eta_{\a\b}$. It means that on the effective background manifold $\bar M$ the coordinate time $u$ is identical to the proper time $\t$ measured on the worldline ${\cal W}$ of the origin of the local coordinates adapted to body B,
\be\la{2n8}
\t=u\;.
\ee 

Post-Newtonian transformation from the global to local coordinates smoothly matches two forms of the background metric, $\bar g_{\a\b}(t,{\bm x})$ and $\bar\rmg_{\a\b}(u,{\bm w})$ on the effective background manifold $\bar M$ in the sense that
\be\label{ww11cc77}
\bar g_{\m\n}(t,{\bm x})= \bar\rmg_{\a\b}(u,{\bm w})\frac{\pd w^\a}{\pd x^\m}\frac{\pd w^\b}{\pd x^\n}\;.
\ee
This should be compared with the law of transformation \eqref{2.6} applied to the full metric $g_{\a\b}$ on spacetime manifold $M$ which includes besides the external part also the internal and internal-external coupling components of the metric tensor perturbations but they are mutually canceled out in \eqref{2.6} leaving only the external terms, thus, converting \eqref{2.6} to \eqref{ww11cc77} without making any additional assumptions about the structure of the effective background manifold $\bar M$. The cancellation of the internal and internal-external components of the metric tensor perturbations in \eqref{2.6} is a manifestation of the {\it effacing} principle \citep{kovl_2008} that excludes the internal structure of body B from the definition of the effective background manifold $\bar M$ used for description of motion of the body \citep{Battista_2017IJMPA}. Compatibility of equations \eqref{2.6} and \eqref{ww11cc77} confirms that the internal and external problems of the relativistic celestial mechanics in $N$-body system are completely decoupled regardless of the structure of the extended bodies and can be extrapolated to compact astrophysical objects like neutron stars and black holes.

In what follows, we will need a matrix of transformation\index{matrix of transformation} taken on the worldline of the origin of the local coordinates,
\be\la{p0c3b5}
\Lambda^\a{}_\b\equiv\Lambda^\a{}_\b(\tau)=\lim_{{\bm x}\rightarrow{\bm x}_\B}\frac{\pd w^\a}{\pd x^\b}\;.
\ee
The components of this matrix can be easily computed from equations of coordinate transformation \eqref{5.12}, \eqref{5.13} and its complete post-Newtonian form is shown in \citep[Section 5.1.3]{kopeikin_2011book}. With an accuracy being sufficient for derivation of the covariant post-Newtonian equations of motion in the present paper, it reads
\ba\la{1as1a}
\Lambda^0{}_0&=&1+\frac12v^2_{\rm B}-\bar U(t,{\bm x}_\B)\;,\\\la{1as1b}
\Lambda^0{}_i&=&-v^i_{\rm B}(1+\frac12v^2_{\rm B})+4\bar U^i(t,{\bm x}_\B)-3v^i_{\rm B}\bar U(t,{\bm x}_\B)\;,\\\la{1as1c}
\Lambda^i{}_0&=&-v^i_{\rm B}\left[1+\frac12v^2_{\rm B}+\bar U(t,{\bm x}_\B)\right]-F^{ij}_\B v^j_{\rm B}\;,\\\la{1as1d}
\Lambda^i{}_j&=&\d^{ij}\left[1+\bar U(t,{\bm x}_\B)\right]+\frac12v^i_{\rm B}v^j_{\rm B}+F^{ij}_\B\;,
\ea
where $F^{ij}_\B$ is the skew-symmetric matrix of the Fermi-Walker precession\index{precession!relativistic} of the spatial axes of the local frame adapted to body B, with respect to the global coordinates -- see \eqref{5.18}. 

We will also need the inverse matrix of transformation between the local and global coordinates taken on the worldline $\cal W$ of the origin of the local coordinates. We shall denote this matrix as
\be\la{p0s8}
\Omega^\a{}_\b\equiv{\Omega}^\a{}_\b(\tau)=\lim_{{\bm w}\rightarrow 0}\frac{\pd x^\a}{\pd w^\b}\;.
\ee
In accordance with the definition of the inverse matrix\index{inverse matrix} we have
\be\la{m9b1z}
\Lambda^\a{}_\b\Omega^\b{}_\g=\d^\a_\g\;,\qquad\qquad \Omega^\a{}_\b\Lambda^\b{}_\g=\d^\a_\g\;.
\ee
 Solving \eqref{m9b1z} with respect to the components of $\Omega^\a{}_\b$, we get
\ba\la{1a2sa}
\Omega^0{}_0&=&1+\frac12v^2_{\rm B}+\bar U(t,{\bm x}_\B)\;,\\\la{1as1b2}
\Omega^0{}_i&=&v^i_{\rm B}(1+\frac12v^2_{\rm B})+F^{ij}_\B v^j_{\rm B}-4\bar U^i(t,{\bm x}_\B)+3v^i_{\rm B}\bar U(t,{\bm x}_\B)\;,\\\la{1as1c2}
\Omega^i{}_0&=&v^i_{\rm B}\left[1+\frac12v^2_{\rm B}+\bar U(t,{\bm x}_\B)\right]\;,\\\la{1as1d2}
\Omega^i{}_j&=&\d^{ij}\left[1-\bar U(t,{\bm x}_\B)\right]+\frac12v^i_{\rm B}v^j_{\rm B}-F^{ij}_\B\;,
\ea
As we shall see below, the matrices $\Lambda^\a{}_\b$ and $\Omega^\a{}_\b$ are instrumental in lifting the geometric objects that are pinned down to the worldline ${\cal W}$ and residing on 3-dimensional hypersurface of constant time $u$ to 4-dimensional effective background manifold $\bar M$.   

In order to arrive to the covariant formulation of the translational and rotational equations of motion we take the equations of motion derived in the local coordinates of body B, and prolong them to the 4-dimensional, covariant form with the help of the transformation matrices and replacing the partial derivatives with the covariant ones. This is in accordance with the Einstein principle of equivalence which establishes a correspondence between spacetime manifold and its tangent space \citep{mtw}. It turns out that, eventually, all direct and inverse transformation matrices cancel out due to \eqref{m9b1z} and the equations acquire a final, covariant 4-dimensional form without any reference to the original coordinate charts that were used in the intermediate transformations. In what follows, we carry out these type of calculations.

\section{The Center-of Mass Worldline as a Perturbed Geodesic on the Effective Background Manifold}\la{o8b4m1}

Our algorithm of derivation of equations of motion defines the center of mass of body B by equating the internal mass-dipole of the body to zero, ${\cal M}^i=0$. The linear momentum, ${\mathfrak{p}}^i$ also vanishes ${\mathfrak{p}}^i=d{\cal M}^i/du=0$, as explained above in text accompanying equation \ref{n5vz1o}. We have shown that these two conditions can be always satisfied by choosing the appropriate value \eqref{q6v4m} of the local acceleration, ${\cal Q}_i$, of the origin of the local coordinates adapted to body B in such a way that the worldline ${\cal W}$ of the origin of the local coordinates coincides with the worldline ${\cal Z}$ of the center of mass of the body. This specific choice of ${\cal Q}_i$ converts the equations of motion of the origin of the local coordinates of body B \eqref{5.8} to the equations of motion of its center of mass in the global coordinates. Below we prove that this equation can be interpreted on the background manifold $\bar M$ as an equation of time-like geodesic of a massive particle with mass, ${\cal M}$, of body B that is perturbed by the force of inertia caused by the local acceleration ${\cal Q}_i$ of the origin of the local coordinates. 

Let us introduce a 4-velocity ${u}^\a$ of the center of mass of body B. In the global coordinates, $x^\a$, the worldline $\cal Z$ of the body's center of mass is described parametrically by $x^0_{\rm B}=t$, and $x^i_{\rm B}(t)$. The 4-velocity is defined by
\be\la{n8c3d}
{u}^\a=\frac{dx^\a_{\rm B}}{d\t}\;,
\ee
where $\t$ is the proper time along the worldline $\cal Z$. The increment $d\t$ of the proper time is related to the increments $dx^\a$ of the global coordinates by equation, 
\be\la{n5b7x}
d\t^2=-\bar{g}_{\a\b}dx^\a dx^\b\;,
\ee
which tells us that the 4-velocity (\ref{n8c3d}) is normalized to unity, ${u}_\a {u}^\a=\bar g_{\a\b}{u}^\a {u}^\b=-1$. In the local coordinates the worldline ${\cal Z}$ is given by equations, $w^\a=(\t,\,w^i=0)$, and the 4-velocity has components $\bar{\rm u}^\a=(1,0,0,0)$. In the global coordinates the components of the 4-velocity are, ${u}^\a=\left(dt/d\t,\,dx^i_{\rm B}/d\t\right)$, which yields 3-dimensional velocity of the body's center of mass, $v^i_\B={u}^i/{u}^0=dx^i_{\rm B}/dt$.  Components of the 4-velocity are transformed from the local to global coordinates in accordance to the transformation equation, ${u}^\a=\Omega^\a{}_\b{\bar{\rm u}}^\b$, which points out that in the global coordinates ${u}^\a=\Omega^\a{}_0$. On the other hand, a covector of 4-velocity obeys the transformation equation, ${u}_\a=\Lambda^\b{}_\a{\bar{\rm u}}_\b$, where ${\bar{\rm u}}_\a=(-1,0,0,0)$ are components of the covector of 4-velocity in the local coordinates. Thus, in the global coordinates ${u}_\a=-\Lambda^0{}_\a$. The above presentation of the components of 4-velocity in terms of the matrices of transformation along with equation \eqref{m9b1z} makes it evident that 4-velocity is subject to two reciprocal conditions of orthogonality,
\be\la{brtv68h}
\Lambda^i{}_\a {u}^\a=0\;, \qquad\qquad {u}_\a\Omega^\a{}_i=0\;.
\ee
Equations \eqref{brtv68h} will be used later on in the procedure of lifting the spatial components of the internal and external multipoles to the covariant form.

In the covariant description of the equations of motion, an extended body B from ${N}$-body system is treated as a particle having mass ${\cal M}$, the mass multipoles ${\cal M}^L$, and the spin multipoles ${\cal S}^L$ attached to the particle, in other words, to the center of mass of the body. This set of the internal multipoles fully characterizes the internal structure of the body. The multipoles, in general, depend on time including the mass of the body which temporal variation (\ref{x1674}) is caused by gravitational coupling of the internal and external multipoles. The mass and spin multipoles are fully determined by their spatial components in the body-adapted local coordinates in terms of integrals from the stress-energy distribution of matter through the solution of the field equations. Covariant generalization of the multipoles from the spatial to spacetime components is provided by the condition of orthogonality of the multipoles to the 4-velocity ${u}^\a$ of the center of mass of the body as explained below in section \ref{n4v7a9}.

We postulate that the covariant definition of the linear momentum of the body is 
\be\label{zowv34as}
{\mathfrak{p}}^\a={\cal M}{u}^\a\;,
\ee 
where ${\mathfrak{p}}^\a$ is a covariant generalization of 3-dimensional linear momentum ${\mathfrak{p}}^i$ of body B introduced in \eqref{b5z0e}. We emphasize that the linear momentum ${\mathfrak{p}}^\a$ may not be reduced to Dixon's linear momentum $p^\a$ in the most general case as comparison of the two definitions \eqref{zowv34as} and \eqref{q13m} show. The Dixon mass, $m$, of body B may be not equal to the post-Newtonian mass, ${\cal M}$, and its {\it dynamic} velocity $\mathfrak{n}^\a$ is not the same as the {\it kinematic} 4-velocity $u^\a$.

We are looking for the covariant translational equations of motion of body B in the following form
\begin{equation}
\label{geodes}
\frac{{\cal D} \mathfrak{p}^\a}{{\cal D}\tau}\equiv {u}^\b\bar{\nabla}_\b {\mathfrak{p}}^\a=\frac{d{\mathfrak{p}}^\a}{d\tau}+\bar\Gamma^\a_{\mu\nu}{\mathfrak{p}}^\mu {u}^\nu=F^{\a},
\end{equation}
where $F^\a$ is a 4-force that causes the worldline $\cal Z$ of the center of mass of the body to deviate from the geodesic worldline of the effective background manifold $\bar M$. We introduce this force to equation (\ref{geodes}) because the body's center of mass experiences a local acceleration ${\cal Q}_i$ given by (\ref{q6v4m}) which means that it is not in the state of a free fall and does not move on geodesic of the background manifold $\bar M$. In order to establish the mathematical form of the force $F^\a$ we split it in two parts,
\be\label{jkev643}
F^\a=F^\a_\perp+F^\a_\parallel\;,
\ee
where $F^\a_\perp$ and $F^\a_\parallel$ are the force components which are orthogonal and parallel to 4-velocity $u^\a$ respectively, $u_\a F^\a_\perp=0$ and $F^\a_\parallel\equiv F u^\a$ where $F$ is yet unknown function of time. It is more convenient to re-write \eqref{geodes} in terms of 4-acceleration $a^\a\equiv {\cal D}{u}^\a/{\cal D\tau}={u}^\b \bar\nabla_\b {u}^\a$,
\be\la{betv67h}
{\cal M}a^\a\equiv{\cal M}\left(\frac{d{u}^\a}{d\tau}+\bar\Gamma^\a_{\mu\nu}{u}^\mu {u}^\nu\right)=F^\a_\perp+\left(F-\dot{\cal M}\right){u}^\a\;,
\ee
where $\dot {\cal M}$ is given in (\ref{x1674}). Multiplying both sides of \eqref{betv67h} with $u^\a$ and taking into account the orthogonality $u_\a a^\a=0$, we get
\be\label{hv3c2z}
F=\dot{\cal M}\;.
\ee

In what follows, it is more convenient to operate with a 4-force per unit mass defined by $f^{\a}\equiv F^\a_\perp/{\cal M}$. After accounting for \eqref{hv3c2z} the equation of motion \eqref{betv67h} is reduced to
\be\la{betv88h}
\frac{d{u}^\a}{d\tau}+\bar\Gamma^\a_{\mu\nu}{u}^\mu {u}^\nu=f^{\a}\;.
\ee
The force $f^\a$ is orthogonal to 4-velocity, ${u}_\a f^\a=0$ as a consequence of its definition. Hence, in the local coordinates of body B the time component of the force vanishes. In the global coordinates the time component of the force $f^\a$ is related to its spatial components as follows, $f_0=-v^i_{\rm B}f_i$. The condition of the orthogonality also yields the contravariant time component of the force in terms of its spatial components,
\be\la{q8h7m}
f^0=-\frac1{\bar g_{00}}\bar g_{ij}v^i_{\rm B}f^j\;.
\ee 
Our task is to prove that the covariant equation of motion (\ref{betv88h}) is exactly the same as the equation of motion (\ref{5.8}) of the center of mass of body B derived in the global coordinates that was obtained by asymptotic matching of the external and internal solutions of the field equations.
To this end we re-parameterize equation (\ref{betv88h}) by coordinate time $t$ instead of the proper time $\tau$, which yields
\ba\la{b4c8d}
a^i_{\rm B}&=&-\bar\G^i_{00}-2\bar\G^i_{0p}v^p_{\rm B}-\bar\G^i_{pq}v^p_{\rm B}v^q_{\rm B}
+\left(\bar\G^0_{00}+2\bar\G^0_{0p}v^p_{\rm B}+\bar\G^0_{pq}v^p_{\rm B}v^q_{\rm B}\right)v^i_{\rm B}+\left(f^i-f^0v^i_{\rm B}\right)\left(\frac{d\tau}{dt}\right)^2\;,
\ea
where $v^i_{\rm B}=dx^i_{\rm B}/dt$ and $a^i_{\rm B}=dv^i_{\rm B}/dt$ are the coordinate velocity and acceleration of the body's center of mass with respect to the global coordinates. 

We calculate the Christoffel symbols, $\bar\G^\a_{\m\n}$, the derivative $d\tau/dt$, substitute them to (\ref{b4c8d}) along with (\ref{q8h7m}), and retain only the Newtonian and post-Newtonian terms. Equation \eqref{b4c8d} takes on the following form
\begin{eqnarray}
  \label{aBi1}
  a_{\rm B}^i & = & \pd^i\bar{U}(t,{\bm x}_\B)-\frac{1}{2}\pd_{tt}\pd^i\bar{\chi}(t,{\bm x}_\B)+4\dot{\bar{U}}^{i}(t,{\bm x}_\B)
   -4v_{\rm B}^j\pd^i\bar{U}^j(t,{\bm x}_\B) -3 v_{\rm B}^i\dot{\bar{U}}(t,{\bm x}_\B)
  -4\bar{U}(t,{\bm x}_\B)\pd^i\bar{U}(t,{\bm x}_\B)\\& +& v_{\rm B}^2 \pd^i\bar{U}(t,{\bm x}_\B) - v_{\rm B}^iv_{\rm B}^j\pd^j\bar{U}(t,{\bm x}_\B)
   +f^i-v_{\rm B}^iv^k_{\rm B}f^k-\left[2\bar{U}(t,{\bm x}_\B)+v_{\rm B}^2\right]f^i\nonumber\;.
\end{eqnarray}
This equation exactly matches translational equation of motion (\ref{5.8}) if we make the following identification of the spatial components $f^i$ of the force per unit mass with the local acceleration ${\cal Q}^i$,
\be\la{c5f6}
f^i\equiv -{\cal Q}^i-\frac12v^i_{\rm B}v^j_{\rm B}{\cal Q}_j+F^{ij}_\B {\cal Q}_j+\bar U(t,{\bm x}_\B){\cal Q}^i\;,
\ee 
By simple inspection we reveal that the right-hand side of the post-Newtonian force (\ref{c5f6}) can be written down in a covariant form 
\be\la{f5t8}
f^\a= -\bar g^{\a\b}\Lambda^{i}{}_\b {\cal Q}_{i}=\bar g^{\a\b}{\cal Q}_\b=-{\cal Q}^\a\;,
\ee
where $\Lambda^{i}{}_\b$ is given above in (\ref{1as1a})-(\ref{1as1d}), and ${\cal Q}_i$ is a vector of 4-acceleration in the local coordinates. The quantity ${\cal Q}_\a=\Lambda^{i}{}_\a {\cal Q}_{i}$ defines the covariant form of the local acceleration in the global coordinates with ${\cal Q}_\a$ being orthogonal to 4-velocity, ${u}^\a {\cal Q}_\a=0$, which is a direct consequence of the condition \eqref{brtv68h}.  Explicit form of ${\cal Q}_i$ in the local coordinates is given in (\ref{q6v4m}) and should be used in (\ref{f5t8}) along with the covariant form of the external multipoles ${\cal Q}_L$, ${\cal C}_L$, ${\cal P}_L$ and the internal multipoles ${\cal M}^L$, $\mathcal{S}^L$ in order to get $f^\a=-\bar g^{a\b}{\cal Q}_\b$. The covariant form of the multipoles is a matter of discussion in next subsection.

\section{Covariant Form of Multipoles}\la{n4v7a9}
\subsection{The Internal Multipoles}

The mathematical procedure that was used in construction of the local coordinates adapted to an extended body B in ${N}$-body system indicates that all type of multipoles are defined at the origin of the local coordinates as the STF Cartesian tensors having only spatial components with their time components being identically nil. It means that the multipoles are projections of 4-dimensional tensors on hyperplane passing through the origin of the local coordinates orthogonal to 4-velocity ${u}^\a$ of the worldline ${\cal Z}$ of the center of mass of the body. The 4-dimensional form of the internal multipoles can be established by making use of the law of transformation from the local to global coordinates,
\be\la{x6v4n}
{\cal M}^{\a_1...\a_l}\equiv\Omega^{\a_1}{}_{i_1}...\Omega^{\a_l}{}_{i_l}{\cal M}^{i_1i_2...i_l}\;,\qquad\qquad
{\cal S}^{\a_1...\a_l}\equiv\Omega^{\a_1}{}_{i_1}...\Omega^{\a_l}{}_{i_l}{\cal S}^{i_1i_2...i_l}\;,
\ee
where the matrix of transformation $\Omega^{\a}{}_{i}$ is given in \eqref{1a2sa}--\eqref{1as1d2}. Transforming 3-dimensional STF Cartesian tensors to 4-dimensional form does not change the property of the tensors to be symmetric and trace-free in the sense that we have for any pair of spacetime (Greek) indices
\be\label{gt4zx}
\bar g_{\a_1\a_2}{\cal M}^{\a_1\a_2...\a_l}=0\;,\qquad\qquad \bar g_{\a_1\a_2}{\cal S}^{\a_1\a_2...\a_l}=0\;.
\ee
The 4-dimensional form \eqref{x6v4n} of the multipoles along with equation \eqref{brtv68h} confirms that the multipoles are orthogonal to 4-velocity, that is
\be\la{m1zz1}
{u}_{\a_1} {\cal M}^{\a_1...\a_l}=0\;,\qquad\qquad
{u}_{\a_1} {\cal S}^{\a_1...\a_l}=0\;,
\ee
and due to the symmetry of the internal multipoles, equation (\ref{m1zz1}) is valid to each index. 

Notice that the matrix of transformation\index{matrix of transformation} (\ref{p0s8}) has been used in making up the contravariant components of the multipoles (\ref{x6v4n}) which are tensors of type $\genfrac[]{0pt}{2}{l}{0}$. Tensor components of the multipoles, ${\cal M}_{\a_1...\a_l}$ and ${\cal S}_{\a_1...\a_l}$, which are of the type $\genfrac[]{0pt}{2}{0}{l}$ are obtained by lowering each index of ${\cal M}^{\a_1...\a_l}$ and ${\cal S}^{\a_1...\a_l}$  respectively with the help of the background metric tensor $\bar g_{\a\b}$. It is worth emphasizing that we have introduced 4-dimensional definitions of the internal multipoles as tensors of type $\genfrac[]{0pt}{2}{l}{0}$ on the ground of transformation equations $\eqref{x6v4n}$ because we defined the spatial components of ${\cal M}^{i_1...i_l}$ and ${\cal S}^{i_1...i_l}$ as integrals \eqref{1.31} and \eqref{1.32} taken from the STF products of the components of 3-dimensional coordinate $w^i$ which behaves as a vector under the linear coordinate transformations. Another reason to use the contravariant components ${\cal M}^{i_1...i_l}$ and ${\cal S}^{i_1...i_l}$ as a starting point for their 4-dimensional prolongation is that the internal multipoles are the coefficients of the Cartesian tensors of type $\genfrac[]{0pt}{2}{l}{0}$  in the Taylor expansions of the gravitational potentials $U_\B(t,{\bm x})$ and $U^i_\B(t,{\bm x})$ with respect to the components of the partial derivatives $\pd_{i_1...i_l} r^{-1}_\B$ which are considered as the STF Cartesian tensors of type $\genfrac[]{0pt}{2}{0}{l}$.   

\subsection{The External Multipoles}\la{ne7c2x9}
The external multipoles, ${\cal Q}_{i_1...i_l}$ and ${\cal C}_{i_1...i_l}$,  have been defined at the origin of the local coordinates of body B by external solutions of the field equations for the metric tensor  and scalar field in such a way that they are purely spatial STF Cartesian tensors of type $\genfrac[]{0pt}{2}{0}{l}$. It means that 4-dimensional tensor extensions of the external multipoles must be orthogonal to 4-velocity of the origin of the local coordinates which is, by construction, identical to 4-velocity $u^\a$ of the worldline ${\cal Z}$ of the center of mass of the body B,
\be\la{m1zz2}
{u}^{\a_1} {\cal Q}_{\a_1\a_2...\a_l}=0\;,\qquad
{u}^{\a_1} {\cal C}_{\a_1\a_2...\a_l}=0\;.
\ee
These orthogonality conditions suggests that the 4-dimensional components of the external multipoles are obtained from their 3-dimensional counterparts by making use of the matrix of transformation (\ref{p0c3b5}) which yields
\be\la{x6v4na}
{\cal Q}_{\a_1...\a_l}\equiv\Lambda^{i_1}{}_{\a_1}...\Lambda^{i_l}{}_{\a_l}{\cal Q}_{i_1...i_l}\;,\qquad\quad
{\cal C}_{\a_1...\a_l}\equiv\Lambda^{i_1}{}_{\a_1}...\Lambda^{i_l}{}_{\a_l}{\cal C}_{i_1...i_l}\;.
\ee
We have used in here the matrix of transformation (\ref{p0c3b5}) because the external multipoles are defined originally as tensor coefficients of the  Taylor expansions of the external potentials $\bar U$, $\bar\Psi$, etc., which are expressed in terms of partial derivatives from these potentials and behave under coordinate transformations like tensors of type $\genfrac[]{0pt}{2}{0}{l}$. 
Definitions \eqref{x6v4na} and the properties of the matrices of transformation suggest that 4-dimensional tensors ${\cal Q}_{\a_1...\a_l}$, ${\cal C}_{\a_1...\a_l}$ and ${\cal P}_{\a_1...\a_l}$ are STF tensors in the sense of \eqref{gt4zx} that is $\bar g^{\a_1\a_2}{\cal Q}_{\a_1...\a_l}=0$, etc. It is known that in general relativity the external multipoles, ${\cal Q}_{i_1...i_l}$ and ${\cal C}_{i_1...i_l}$ are defined in the local coordinates by partial derivatives of the Riemann tensor, $\bar R^\a{}_{\m\b\n}$, of the background metric (\ref{k8b2d}) taken at the origin of the local coordinates \citep{th_1985,1986PhRvD..34.3617S,zhang_1986PhRvD,poisson_2011}. 

As we show below, the 4-dimensional tensor formulation of the external multipoles is achieved by contracting the Riemann tensor with vectors of 4-velocity, ${u}^\a$, and taking the covariant derivatives $\bar\nabla_\a$ projected on the hyperplane being orthogonal to the 4-velocity. The projection is fulfilled with the help of the operator of projection,
\be\la{n6g0s}
\pi^{\alpha}_{\beta}\equiv \delta^{\alpha}_{\beta}+{u}^{\alpha}{u}_{\beta}\;, \qquad\quad\pi^{\a\b}=\bar g^{\a\b}+{u}^\a {u}^\b\;,\qquad\quad\pi_{\a\b}=\bar g_{\a\b}+{u}_\a {u}_\b\;,  
\ee
The operator of projection satisfies the following relations: $\pi^\alpha_\g\pi^\g_\beta=\pi^\alpha_\beta$, $\pi^{\a\b}=\bar g^{\a\g}\pi^\b_\g$, $\pi_{\a\b}=\bar g_{\a\g}\pi^\g_\b$, and $\pi^\a_\a=3$. The latter property points out that $\pi^{\alpha}_{\beta}$ has only three algebraically-independent components which are reduced to the Kronecker symbol when $\pi^\a_\b$ is computed in the local coordinates of body B, that is in the local coordinates $\pi^0_0=0\;,\pi^i_0=\pi^0_i=0\;,\pi^i_j=\d^i_j$. In other words, the projection operator is a 3-dimensional Kronecker symbol $\d^i_j$ lifted up to 4-dimensional effective background manifold $\bar M$. We notice that the operator of the projection has some additional algebraic properties. Namely,
\be\la{nrv7b3}
\pi^\a_\b\Lambda^i{}_\a=\Lambda^i{}_\b\;,\qquad\qquad \pi^\b_\a\Omega^\a{}_i=\Omega^\b{}_i\;,
\ee
that are in accordance with the condition of orthogonality \eqref{brtv68h}. They point out that $\pi^{\alpha}_{\beta}$ can be also represented as a product of two reciprocal transformation matrices,
\be
\pi^{\alpha}_{\beta}=\Omega^\a{}_i\Lambda^i{}_\b\;.
\ee

The projection operator is required to extend 3-dimensional spatial derivatives of geometric objects to their 4-dimensional counterparts. Indeed, in the local coordinates the external multipoles are purely spatial Cartesian tensors which are expressed in terms of the partial spatial derivatives of the external perturbations of the metric tensor. It means that the extension of a spatial partial derivative to its 4-dimensional form must preserve its orthogonality to the 4-velocity ${u}^\a$ of the worldline ${\cal Z}$ which is achieved by coupling the spatial derivatives with the projection operator. Covariant form of 3-dimensional STF multipoles being orthogonal to 4-velocity $u^\a$ is obtained from the standard definition of the Cartesian STF tensors \eqref{stfformula} by extending 3-dimensional Kronecker symbol and other 3-tensors to 4-dimensional form by making use of the Einstein equivalence principle, 
\be\label{stf4form}
T_{<\a_1...\a_l>}\equiv\sum_{n=0}^{[l/2]}\frac{(-1)^n}{2^nn!}\frac{l!}{(l-2n)!}\frac{(2l-2n-1)!!}{(2l-1)!!}\pi_{(\a_1\a_2...}\pi_{\a_{2n-1}\a_{2n}}T_{(\a_{2n+1}...\a_l)\b_1\g_1...\b_n\g_n)}\pi^{\b_1\g_1}...\pi^{\b_n\g_n}\;.
\ee 
We also notice that the projection operator can be effectively used to rise and to lower 4-dimensional (Greek) indices of the internal and external multipoles like the metric tensor $\bar g_{\a\b}$. This is because all multipoles are orthogonal to 4-velocity ${u}^\a$. Thus, for example, ${\cal Q}_{\a\b}\bar g^{\b\g}={\cal Q}_{\a\b}\pi^{\b\g}={\cal Q}_\a{}^\g$, etc.

The external multipoles ${\cal Q}_{\a_1...\a_l}$ and ${\cal C}_{\a_1...\a_l}$ are directly connected to the Riemann tensor of the background manifold $\bar M$ and its covariant derivatives. In order to establish this connection we work in the local coordinates and employ covariant definition \eqref{kk33cc25} to compute the Riemann tensor on the effective background manifold $\bar M$ of body B, 
\be\label{kk33xc}
\bar\R_{\a\b\m\n}=\frac12\left(\pd_{\a\n} \bar \rmg_{\b\m}+\pd_{\b\m} \bar \rmg_{\a\n}-\pd_{\b\n} \bar \rmg_{\a\m}-\pd_{\a\m} \bar \rmg_{\b\n}\right)+ \bar \rmg_{\r\s}\left(\bar\Upgamma^\r_{\a\n}\bar\Upgamma^\s_{\b\m}-\bar\Upgamma^\r_{\a\m}\bar\Upgamma^\s_{\b\n}\right)\;.
\ee
where the metric tensor in the local coordinates is given by equation \eqref{k8b2d}.  
The products of the Christoffel symbols entering the Riemann tensor at the post-Newtonian level of approximation require to know the following components of the Christoffel symbols
\be\label{kk22ss}
\bar\Upgamma^i_{00}=\bar\Upgamma^0_{0i}=-\frac12 \pd_i\rmh_{00}^{E}\;,\qquad\qquad \bar\Upgamma^i_{jk}= \frac12\left(\pd_j\rmh_{ik}^{E}+\pd_k\rmh_{ik}^{E}-\pd_i\rmh_{jk}^{E}\right)\;.
\ee
Substituting \eqref{k8b2d} and \eqref{kk22ss} to \eqref{kk33xc} and taking into account all post-Newtonian terms we get the STF part of the Riemann tensor component $[\bar \R_{0i0j}]^{\rm STF}\equiv \bar \R_{0<i|0|j>}$ in the following form \footnote{Vertical bars around index indicate that the index is excluded from STF projection.},
\ba\la{riem634}
 \left[\bar\R_{0i0j}\right]^{\rm STF}&=&-D_{<ij>}+3D_{<i}D_{j>}+2DD_{<ij>}
 \\\nonumber
&+&\sum_{l=0}^\infty\frac1{l!}\left[\frac{2(l-1)}{(2l+5)(l+2)}\ddot {\cal Q}_{L<i}w_{j>L}
-\frac{l+7}{2(2l+7)(l+3)}\ddot {\cal Q}_{<ij>L}w^{L}w^2
+\frac{1}{l+2}\varepsilon_{pq<i}\dot {\cal C}_{j>pL}w^{qL}\right]\;,
\ea
where we have discarded all terms of the post-post-Newtonian order and
introduced the shorthand notations 
\be
D\equiv \sum_{k=1}^\infty \frac1{k!}{\cal Q}_{K}(u)w^K\;,\qquad\qquad 
D_{i_1...i_l}\equiv\partial_{i_1...i_l}D=\sum_{k=0}^\infty \frac1{k!}{\cal Q}_{i_1...i_lK}(u)w^K\;.
\ee
Notice that at the origin of the local coordinates where $w^i=0$, we have $D=0$ and $D_{i_1...i_l}={\cal Q}_{i_1...i_l}$. Therefore, at the origin of the local coordinates, that is on the worldline ${\cal Z}$, the value of the STF Riemann tensor \eqref{riem634} is simplified to
\be\la{evy53c}
\left[\bar \R_{0i0j}\right]^{\rm STF}_{\cal Z}=-{\cal Q}_{<ij>}+3{\cal Q}_{<i}{\cal Q}_{j>}\;.
\ee
This relationship establishes the connection between the external mass quadrupole ${\cal Q}_{ij}$ and the STF Riemann tensor. The reader should notice that \eqref{evy53c} includes terms depending on acceleration ${\cal Q}_i$ of the worldline of the center of mass of body B. This may look strange as the curvature of spacetime (the Riemann tensor) does not depend on the choice of the worldline of the local coordinates.  Indeed, it can be verified that the acceleration-dependent terms in \eqref{evy53c} are mutually canceled out with the similar terms coming out of the explicit expression for $X^L$ term in ${\cal Q}_{ij}$ -- see \eqref{5.9} and \eqref{1q4d}.  

Relationship between the STF covariant derivative of $l$-th order from the Riemann tensor and the external gravitoelectric multipole of the same order is derived by taking covariant derivatives $l$ times from both sides of \eqref{riem634}. Covariant derivative of the order $l$ from the Riemann tensor is a linear operator on the background manifold $\bar M$ that involves the products of the Christoffel symbols and the covariant derivatives of the order $l-1$ from the Riemann tensor. They can be calculated by iterations starting from $l=1$. Straightforward but tedious calculation shows that at the post-Newtonian level of approximation the covariant derivative of the order $l-2$ combined with the Riemann tensor to STF tensor of the order $l$, reads,
\be\la{u5b1x}
\left[\bar\nabla_{i_1...i_{l-2}}\bar{\R}_{0i_{l-1}0i_l}\right]^{\rm STF}=
\left[\pd_{i_1...i_{l-2}}\bar{\R}_{0i_{l-1}0i_l}\right]^{\rm STF}
+2\sum_{k=0}^{l-3}(k+1)\pd_{<i_1...i_{l-k-3}}\left[D_{i_{l-k-2}...i_{l-1}}D_{i_l>}\right]\;.
\ee
Applying the Leibniz rule of differentiation to the product of two functions \citep[Equation 0.42]{gradryzh}  standing in the right hand-side of \eqref{u5b1x}, we obtain a more simple expression,
\ba
\left[\bar\nabla_{i_1...i_{l-2}}\bar{\R}_{0i_{l-1}0i_l}\right]^{\rm STF}&=&
\left[\pd_{i_1...i_{l-2}}\bar{\R}_{0i_{l-1}0i_l}\right]^{\rm STF}
+2\sum_{k=0}^{l-3}\sum_{s=0}^{k}\frac{(l-k-2)k!}{s!(k-s)!}D_{<i_1...i_{s+1}}D_{i_{s+2}...i_l>}\;.
\ea
The $l-2$-th order partial derivatives from terms $D_{<i}D_{j>}$, $DD_{<ij>}$, etc., entering $\left[\pd_{i_1...i_{l-2}}\bar \R_{0i_{l-1}0i_l}\right]^{\rm STF}$, are also calculated with the help of the Leibniz rule, yielding 
\ba
\pd_{<i_1...i_{l-2}}\left[D_{i_{l-1}}D_{i_l>}\right]&=&\sum_{k=0}^{l-2}\frac{(l-2)!}{k!(l-k-2)!}D_{<i_1...i_{k+1}}D_{i_{k+2}...i_{l}>}\;,\\
\pd_{<i_1...i_{l-2}}\left[D_{i_{l-1}i_l>}D\right]&=&
\sum_{k=1}^{l-2}\frac{(l-2)!}{k!(l-k-2)!}D_{<i_1...i_k}D_{i_{k+1}...i_l>}\;.
\ea

Actually, we need the covariant derivatives of the STF part of the Riemann tensor only at the origin of the local coordinates adapted to body B. Therefore, after taking the STF covariant derivatives from the Riemann tensor we take the value of the local spatial coordinates $w^i=0$, which eliminates all terms depending on the time derivatives of the external multipoles in the right hand side of \eqref{riem634} for the STF part of the Riemann tensor. Hence, the STF covariant derivative of the Riemann tensor taken on the worldline of the center of mass of body B reads,
\ba\la{vx4a6c}
\left[\bar\nabla_{i_1...i_{l-2}}\bar{\R}_{0i_{l-1}0i_l}\right]^{\rm STF}_{\cal Z}&=&
-{\cal Q}_{<i_1...i_l>}+
3\sum_{k=0}^{l-2}\frac{(l-2)!}{k!(l-k-2)!}{\cal Q}_{<i_1...i_{k+1}}{\cal Q}_{i_{k+2}...i_{l}>}\\\nonumber
&+&2\bigg[\sum_{k=1}^{l-2}\frac{(l-2)!}{k!(l-k-2)!}{\cal Q}_{<i_1...i_k}{\cal Q}_{i_{k+1}...i_{l}>}+\sum_{k=0}^{l-3}\sum_{s=0}^{k}\frac{(l-k-2)k!}{s!(k-s)!}{\cal Q}_{<i_1...i_{s+1}}{\cal Q}_{i_{s+2}...i_l>}\bigg]\;.
\ea
 
It is rather straightforward now to convert \eqref{vx4a6c} to 4-dimensional form valid in arbitrary coordinates on the effective background manifold $\bar M$ by making use of the transformation matrices and the operator of projection as it was explained above. We introduce a new notation for the covariant STF derivative of the Riemann tensor taken on the worldline ${\cal Z}$,
\be
  \label{sd11}
{\cal E}_{\a_1...\a_l}  \equiv {\pi}^{\beta_1}_{<\alpha_1}\pi^{\b_2}_{\a_2}.... {\pi}^{\beta_l}_{\alpha_l>}\left[\bar\nabla_{\beta_1...\beta_{l-2}}\bar R_{\mu\beta_{l-1}\beta_l\nu} {u}^{\mu}{u}^{\nu}\right]^{\rm STF}_{\cal Z}\;,
\ee
and use it for transformation of \eqref{vx4a6c} to arbitrary coordinates. It yields a covariant expression for the external gravitoelectric multipoles ${\cal Q}_{\a_1...\a_l}$ in terms of the STF covariant derivatives from the Riemann tensor,
\begin{eqnarray}
\label{we69}
{\cal Q}_{\a_1...\a_l} & = &  {\cal E}_{<\a_1...\a_l>}
+3\sum_{k=0}^{l-2}\frac{(l-2)!}{k!(l-k-2)!}{\cal E}_{<\a_1...\a_{k+1}}{\cal E}_{\a_{k+2}...\a_{l}>}\\\nonumber
&+&2\bigg[\sum_{k=1}^{l-2}\frac{(l-2)!}{k!(l-k-2)!}{\cal E}_{<\a_1...\a_k}{\cal E}_{\a_{k+1}...\a_{l}>}+\sum_{k=0}^{l-3}\sum_{s=0}^{k}\frac{(l-k-2)k!}{s!(k-s)!}{\cal E}_{<\a_1...\a_{s+1}}{\cal E}_{\a_{s+2}...\a_l>}\bigg]\;,
\ea
where 
we have made identification: ${\cal E}_a\equiv{\cal Q}_\a$. 
  
Similar, but less tedious procedure allows us to calculate 4-dimensional form of the external gravitomagnetic multipoles ${\cal C}_{\a_1...\a_l}$ in terms of the STF covariant derivative of the Riemann tensor. We get,
\be
\label{we70}
{\cal C}_{\a_1...\a_l}  \equiv 
 {\pi}^{\beta_1}_{<\alpha_1}\pi^{\b_2}_{\a_2}...{\pi}^{\b_l}_{\alpha_l>}\left[\bar\nabla_{\beta_1...\beta_{l-2}}\bar R_{\s\m\n\beta_{l-1}}\varepsilon_{\b_l}{}^{\s\m}{u}^\n\right]^{\rm STF}_{\cal Z}\;.
\ee
where we have utilized 3-dimensional covariant tensor of Levi-Civita, $\varepsilon_{\a\beta\gamma}$, which is equivalent to projection of 4-dimensional, fully-antisymmetric Levi-Civita symbol $E_{\a\mu\nu\r}=E^{\a\mu\nu\r}$ \citep[\S 3.5]{mtw} on the hyperplane being orthogonal to 4-velocity ${u}^\a$. By definition \citep[\S 8.4]{mtw},
\be\label{vareps67}
\varepsilon_{\a\beta\gamma}\equiv \sqrt{-\bar g}u^\m E_{\m\a\b\g}\;,\qquad\qquad \varepsilon^{\a\beta\gamma}
\equiv -\frac1{\sqrt{-\bar g}}u_\m  E^{\m\a\b\g}\;,
\ee
and we can easily check by inspection that $\varepsilon_{\a\beta\gamma}$ is lying in the hyperplane being orthogonal to 4-velocity $u^\a$ of the center of mass of body B, $u^\a\varepsilon_{\a\beta\gamma}=0$.

Covariant 4-dimensional prolongations of the external multipoles allow us to transform products of the multipoles given in the local coordinates to their covariant counterparts, for example, ${\cal Q}_{L}{\cal M}^L\equiv {\cal Q}_{i_1...i_l}{\cal M}^{i_1...i_l}={\cal Q}_{\a_1...\a_l}{\cal M}^{\a_1...\a_l}$, etc. In all such products the matrices of transformation cancel out giving rise to covariant expressions being independent of a particular choice of coordinates. 

\section{Post-Newtonian Covariant Equations of Motion}\la{nxv34xd}
\subsection{Translational Equations for Linear Momentum}\la{n3cz52s}

A generic form of the covariant translational equations of motion have been formulated in \eqref{geodes}--\eqref{hv3c2z}. It reads
\begin{eqnarray}
  \label{ubpu;b}
 \frac{{\cal D} {\mathfrak{p}}^\m}{{\cal D}\tau}&=&F^\mu_\perp+\dot{\cal M}u^\mu \;,
\end{eqnarray}
where force $F^\a_\perp={\cal M}f^\a=-{\cal M}{\cal Q}^\a$ 
and the second term in the right hand side of \eqref{ubpu;b} is due to the non-conservation of mass (\ref{x1674}) having the following covariant form 
\ba\la{x1674qq}
\dot {\cal M}&=&
-\sum\limits_{l=2}^\infty\frac1{(l-1)!}{\cal Q}_{\a_1...\a_l}\frac{{\cal D}_{\rm F}{\cal M}^{\a_1...\a_l}}{{\cal D}\tau}-\sum\limits_{l=2}^\infty\frac{l+1}{l!}{\cal M}^{\a_1...\a_l}\frac{{\cal D}_{\rm F}{\cal Q}_{\a_1...\a_l}}{{\cal D}\tau}\;,
\ea
where we have used the center-of-mass condition ${\cal M}^\a=0$.
Equation \eqref{x1674qq} includes the Fermi-Walker derivative ${\cal D}_{\rm F}/{\cal D}\tau$ of the multipole moments which is a covariant generalization of the total time derivative in the body-adapted local coordinates. The Fermi-Walker derivative is explained in more detail at the end of this section -- see equation \eqref{p2b8r3}.

Gravitational force $F^\mu_\perp$ in the right hand side of \eqref{ubpu;b} is the 4-dimensional extension of 3-dimensional force \eqref{f5t8} with the local 4-acceleration ${\cal Q}_i$ defined in \eqref{q6v4m}. It can be represented in the following form,
\be\la{g6xc2c7}
F^\mu_\perp={F}^\m_{\cal Q}+{F}^\m_{\cal C}\;,
\ee
and describes the gravitational interaction between the internal multipoles of body B and the external gravitoelectric and gravitomagnetic multipoles. We have,
\begin{eqnarray}
  \label{sd15a}
  F^{\mu}_{\cal Q} & = & \sum_{l=2}^{\infty}\frac{1}{l!}\bar g^{\mu\nu}{\cal Q}_{\nu\a_1...\a_l}{\cal M}^{\a_1...\a_l}
   -\sum_{l=2}^{\infty}\frac{l^2+l+4}{(l+1)!}{\cal Q}_{\a_1...\a_l}\frac{{\cal D}^2_{\rm F}{\cal M}^{\mu\a_1...\a_l}}{{\cal D}\tau^2}\\
  &-& \sum_{l=2}^{\infty}\frac{2l+1}{(l+1)!}\left(\frac{l^2+2l+5}{l+1}\frac{{\cal D}_{\rm F}{\cal Q}_{\a_1...\a_l}}{{\cal D}\tau}\frac{{\cal D}_{\rm F}{\cal M}^{\mu\a_1...\a_l}}{{\cal D}\tau}
   +\frac{l^2+3l+6}{2l+3}{\cal M}^{\mu\a_1...\a_l}\frac{{\cal D}^2_{\rm F}{\cal Q}_{\a_1...\a_l}}{{\cal D}\tau^2}\right)\nonumber\\
 &+&4\sum_{l=1}^{\infty}\frac{l+1}{(l+2)!}\varepsilon^{\mu\rho}{}_{\sigma}\left({\cal Q}_{\rho\a_1...\a_l}\frac{{\cal D}_{\rm F}{\mathcal{S}}^{\sigma\a_1...\a_l}}{{\cal D}\tau}
 +\frac{l+1}{l+2}\mathcal{S}^{\sigma\a_1...\a_l}\frac{{\cal D}_{\rm F} {\cal Q}_{\rho\a_1...\a_l}}{{\cal D}\tau}\right)\nonumber\\
&+&\varepsilon^{\mu\rho}{}_{\sigma}\left(2{\cal Q}_{\rho}\frac{{\cal D}_{\rm F}{\mathcal{S}}^{\sigma}}{{\cal D}\tau}
 +\mathcal{S}^{\sigma}\frac{{\cal D}_{\rm F}{\cal Q}_\rho}{{\cal D}\tau}\right)-3\frac{{\cal D}^2_{\rm F}\left({\cal M}^{\mu\a}{\cal Q}_\a\right)}{{\cal D}\tau^2}\;,\nonumber\\ 
  \label{sd16}
  F^{\mu}_{\cal C}     &= & \sum_{l=1}^{\infty}\frac{l}{(l+1)!}\bar g^{\mu\nu}{\cal C}_{\nu\a_1...\a_l}\mathcal{S}^{\a_1...\a_l}
  -\sum_{l=1}^{\infty}\frac{1}{(l+1)!}\varepsilon^{\mu\rho}{}_\sigma\left[{\cal C}_{\rho\a_1...\a_l}\frac{{\cal D}_{\rm F}{\cal M}^{\sigma\a_1...\a_l}}{{\cal D}\tau}+\frac{l+1}{l+2}{\cal M}^{\sigma\a_1...\a_l}\frac{{\cal D}_{\rm F}{\cal C}_{\rho\a_1...\a_l}}{{\cal D}\tau} 
  \right],
\end{eqnarray}
where again the condition ${\cal M}^\a=0$ have been used.

Time derivatives of the internal and external multipoles of body B in the local coordinates are taken at the fixed value of the spatial coordinates, $w^i=0$, that is at the origin of the body-adapted local coordinates which coincides with the center of mass of body B. The multipoles are STF Cartesian tensors which are orthogonal to 4-velocity of worldline ${\cal Z}$ of the center of mass of the body. This worldline is not geodesic on the effective background manifold $\bar M$ but accelerates with the local acceleration $a^\a=-Q^\a$. Therefore, the time derivative of the multipoles   
corresponds to the Fermi-Walker covariant derivative -- denoted as ${\cal D}_{\rm F}/{\cal D}\tau$ -- on the background manifold $\bar M$\index{manifold!background} taken along the direction of the 4-velocity vector ${u}^\a$ with accounting for the Fermi-Walker transport \citep[Chapter 1, \S4]{syngebook}. For example, the first time derivative taken from 3-dimensional internal multipole $\dot{\cal M}^L\equiv\dot{\cal M}^{i_1i_2...i_l}$ in the local coordinates, is mapped to the 4-dimensional Fermi-Walker covariant derivative as follows, 
\be\la{p2b8r3}
\dot{\cal M}^L\mapsto\frac{{\cal D}_{\rm F}{\cal M}^{\a_1\a_2...\a_l}}{{\cal D}\tau}\equiv\frac{{\cal D}{\cal M}^{\a_1\a_2...\a_l}}{{\cal D}\tau}-la_\b{u}^{<\a_1} {\cal M}^{\a_2...\a_l>\b}\;,
\ee
where ${\cal D}{\cal M}^{<\a_1\a_2...\a_l>}/{\cal D}\tau\equiv {u}^\beta\bar\nabla_\b{\cal M}^{<\a_1\a_2...\a_l>}$ is a standard covariant derivative of tensor ${\cal M}^{<\a_1\a_2...\a_l>}$ along 4-vector $u^\a$, and $a^\a={\cal D}u^\a/{\cal D}\tau=-{\cal Q}^\a$ is 4-acceleration of the origin of the local coordinates adapted to body B -- see equation \eqref{f5t8}. In a similar way, the second time derivative from 3-dimensional internal multipole, $\ddot{\cal M}^L\equiv\ddot{\cal M}^{i_1i_2...i_l}$, taken in the local coordinates, can be mapped to the 4-dimensional Fermi-Walker covariant derivative of the second order by consecutively applying the rule \eqref{p2b8r3} two times,
\ba\la{huc4}
\ddot{\cal M}^L\mapsto\frac{{\cal D}^2_{\rm F} {\cal M}^{\a_1\a_2...\a_l}}{{\cal D}\tau^2}&\equiv& \frac{{\cal D}^2 {\cal M}^{\a_1\a_2...\a_l}}{{\cal D}\tau^2}
-2la_\b{u}^{<\a_1}\frac{{\cal D} {\cal M}^{\a_2...\a_l>\b}}{{\cal D}\tau}\\\nonumber
&-&l\frac{{\cal D}a_\b}{{\cal D}\tau}{u}^{<\a_1} {\cal M}^{\a_2...\a_l>\b}
+la_\b{\cal Q}^{<\a_1} {\cal M}^{\a_2...\a_l>\b}+l^2a_\b{\cal Q}_\g{u}^{<\a_1}{u}^{\a_2} {\cal M}^{\a_3...\a_l>\b\g}
\;,
\ea
where ${\cal D}a^\a/{\cal D}\tau={u}^\b\bar\nabla_\b a^\a$ is the covariant derivative of the 4-acceleration of the origin of the local frame taken along the direction of its 4-velocity. We compare our covariant equations \eqref{ubpu;b}--\eqref{sd16} of translational motion with the corresponding Dixon's equation \eqref{q15ms} in section \ref{ndj45xc}.

\subsection{Rotational Equations for Spin}\label{mev254z4}
Covariant rotational equations of motion generalize 3-dimensional form \eqref{spin-10} of the rotational equations for spin of body B which is a member of $N$-body system. Spin is a vector that is orthogonal to 4-velocity of the worldline ${\cal Z}$ of the center of mass of body B. It is carried out along this worldline according to the Fermi-Walker transportation rule. The covariant form of \eqref{spin-10} is based on the Fermi-Walker derivative, and reads 
\be\label{ui4vsr}
\frac{{\cal D}_{\rm F}{\cal S}^\mu}{{\cal D}\tau}={\cal T}^\mu\;,
\ee
or, after making use of definition \eqref{p2b8r3} of the Fermi-Walker derivative, more explicitly,
\be\label{ui6s41}
\frac{{\cal D}{\cal S}^\mu}{{\cal D}\tau}={\cal T}^\mu-\left({\cal S}^\b {\cal Q}_\b\right) {u}^{\mu}\;,
\ee
where the second term in the right hand side is due to the fact that the Fermi-Walker transport is executed along the accelerated worldline ${\cal Z}$ of the center of mass of body B, the torque ${\cal T}^\mu$ is a covariant generalizations of 3-torque \eqref{spin-5}, 
\ba\label{ac5s03v}
{\cal T}^\mu&=&
-\varepsilon^{\mu\rho}{}_\sigma\sum_{l=1}^{\infty}\frac1{l!}\left({\cal Q}_{\rho\a_1...\a_l}{\cal M}^{\sigma\a_1...\a_l} 
+\frac{l+1}{l+2}{\cal C}_{\rho\a_1...\a_l}{\cal S}^{\sigma\a_1...\a_l}\right)\;,
\ea
where the external multipole moments ${\cal Q}_{\a_1...\a_l}$ and ${\cal C}_{\a_1...\a_l}$ are expressed in terms of the Riemann tensor of the background manifold $\bar M$ in accordance with equations \eqref{we69} and \eqref{we70} respectively. Comparison of our spin evolution equation \eqref{ui6s41} of body B with corresponding Dixon's equation \eqref{q16mc} will be done in section \ref{iopn3e4}.

\section{Comparison of the Dixon Multipoles with the Blanchet-Damour Multipoles}\label{appndxon}
\subsection{Algebraic Properties of the Dixon multipoles}
Before comparing our covariant equations of motion \eqref{ubpu;b}, \eqref{ui6s41} with analogous equations \eqref{q15ms}, \eqref{q16mc} derived by \citet{dixon_1979} in the MPD formalism, we need to establish the correspondence between the Dixon multipole moments $I^{\a_1...\a_l\m\n}$ and the STF mass and spin multipoles ${\cal M}^{\a_1...\a_l}$ and ${\cal S}^{\a_1...\a_l}$ that have been introduced by \citet{bld1986}, and are used in the present paper. 
We, first, discuss the algebraic properties of the Dixon multipoles in more detail. 

\citet{dixon_1979} has defined internal multipoles of an extended body B in the normal Riemann coordinates, $X^\a$, by means of a tensor integral \eqref{q12m} which we repeat for the reader convenience,
\be\label{q12ttt} 
I^{\a_1...\a_l\m\n}(z)=\int X^{\a_1}...X^{\a_l}\hat T^{\m\n}(z,X)\sqrt{-\bar g(z)}DX\;,\qquad\qquad (l\ge 2)
\ee
where $\hat T^{\m\n}$ is the stress-energy {\it skeleton} of the body, the integration is performed in the tangent 4-dimensional space to the effective background manifold $\bar M$ at point $z$ taken on a reference worldline ${\cal Z}$, and the volume element of integration $DX=dX^0\wedge dX^1\wedge dX^2\wedge dX^3$. The reason for appearance of the {skeleton} $\hat T^{\m\n}$ in \eqref{q12ttt} instead of the regular stress-energy tensor $T^{\m\n}$ was to incorporate the self-field effects of gravitational field of the body to the definition of the higher-order multipoles \footnote{The influence of the self-field effects on multipoles was studied by \citet{thor} and \citet{bld,dyr2} with different techniques.}. According to \citep{dixon_1979}, the skeleton $\hat T^{\m\n}$ is a certain distribution \citep{shilov_1968} defined on the tangent bundle to the background spacetime manifold $\bar M$ at each point of worldline ${\cal Z}$ in such a way that it contains a complete information about the body but is entirely independent of the geometry of the surrounding spacetime to which the body is embedded. The skeleton is lying on the hyperplane made out of vectors $X^\a$ which are orthogonal to the vector of dynamic velocity ${\mathfrak n}^\a$. It gives the following constraint \citep[Equation (91)]{dixon_1979}, 
\be\label{2x4a3z11} 
({\mathfrak n}_\a X^\a)X^{[\l}\hat T^{\m][\n}X^{\s]}=0\;,
\ee
which points out that the skeleton distribution is concentrated on the hyperplane ${\mathfrak n}_\a X^\a=0$.

Definition (\ref{q12ttt}) suggests that the Dixon multipole moments have the following symmetries,
\be\la{o39v3j}
I^{\a_1...\a_l\m\n}=I^{(\a_1...\a_l)(\m\n)}\;,
\ee
where the round parentheses around the tensor indices denote a full symmetrization\index{symmetrization}. In addition to \eqref{o39v3j} there are more symmetries of the Dixon multipoles due to the one-to-one mapping of the microscopic equation of motion \eqref{wk10} to a similar equation for the stress-energy {skeleton} in the normal Riemann coordinates \citep{dixon_1979} 
\be\label{on3c45}
\pd_{\n} \hat T^{\m\n}(z,X)=0\;. 
\ee
Multiplying \eqref{on3c45} with $X^{\a_1}...X^{\a_l}X^{\a_{l+1}}$, integrating over 4-dimensional volume and taking into account that $\hat T^{\m\n}$ vanishes outside hyperplane ${\mathfrak n}_\a X^\a=0$, yields \citep[Equation 143]{dixon_1979},
\be\label{gr7xews}
I^{(\a_1...\a_l\m)\n}=0\;,
\ee
and a similar relation holds after exchanging indices $\m$ and $\n$ due to symmetry \eqref{o39v3j}.
The number of algebraically independent components of $I^{\a_1...\a_l\m\n}$ obeying \eqref{o39v3j} is $N_1(l)=C^{l+3}_3\times C^5_3$ where $C^p_q=\frac{p!}{q!(p-q)!}$ is a binomial coefficient. Constraints \eqref{gr7xews} reduce the number of the algebraically independent components of the multipoles $I^{\a_1...\a_l\m\n}$ by $N_2(l)=C^{l+4}_3\times C^4_3$ making the number of linearly independent components of $I^{\a_1...\a_l\m\n}$ equal to
$N_3(l)=N_1(l)-N_2(l)=(l+3)(l+2)(l-1)$.

The multipoles $I^{\a_1...\a_l\m\n}$ are coupled to the Riemann tensor\index{Riemann tensor} $\bar R^\a{}_{\m\b\n}$ characterizing the curvature\index{curvature} of the effective background spacetime. Therefore, they can be replaced with a more suitable set of {\it reduced} moments $J^{\a_1...\a_l\l\m\n\r}$ which are defined by the following formulas \citep{dixon_1979,dixon_1973GReGr}
\be\la{wk12dcr}
J^{\a_1...\a_p\lambda\m\s\n}\equiv I^{\a_1...\a_p[\lambda[\s\m]\n]}\;,
\ee
where the square parentheses around the tensor indices denote a full anti-symmetrization\index{anti-symmetrization}, and the nested square brackets in (\ref{wk12dcr}) denote the anti-simmetrization on pairs of indices $[\l,\m]$ and $[\n,\r]$ independently. Definition \eqref{wk12dcr} tells us that
tensor $J^{\a_1...\a_p\lambda\m\s\n}$ is fully symmetric with respect to the first $p$ indices and is skew-symmetric with respect to the pairs of indices $\lambda,\mu$ and $\s,\n$,
\be
J^{\a_1...\a_p\lambda\m\s\n}=J^{(\a_1...\a_p)[\lambda\m][\s\n]}\;.
\ee  
Among other properties of $J^{\a_1...\a_p\lambda\m\s\n}$ we have
\be
J^{\a_1...\a_p\l[\m\s\n]}=0\;,\qquad\qquad
J^{\a_1...[\a_p\l\m]\s\n}=0\;,
\ee
which are consequences of the definition \eqref{wk12dcr}, and 
\be \label{kerxv34}
{\mathfrak n}_{\a_1} J^{\a_1...\a_p\l\m\s\n}=0\;,
\ee
that is the condition of orthogonality following from the constraint \eqref{2x4a3z11}.

Equation (\ref{wk12dcr}) can be transformed to another form. For this we write down the anti-symmetric part of \eqref{wk12dcr} explicitly as a combination of four terms, change notations of indices $\{\a_1...\a_p\m\n\}\rightarrow\{\a_1...\a_{l-2}\a_{l-1}\a_l\}$, and make a full symmetrization with respect to the set of indices $\{\a_1...\a_l\}$. It gives,
\be\label{bwc3aj}
J^{(\a_1...\a_{l-1}|\m|\a_l)\n}=\frac14\left[I^{(\a_1...\a_{l-1}\a_l)\m\n}-I^{(\a_1...\a_{l-2}|\m|\a_{l-1}\a_l)\n}-I^{(\a_1...\a_{l-2}\a_{l-1}|\n\m|\a_l)}+I^{(\a_1...\a_{l-2}|\m\n|\a_{l-1}\a_l)}\right]\;,
\ee
where the indices enclosed to vertical bars are excluded from symmetrization. Remembering that each of the $I$ moments is separately symmetric with respect to the first $l$ and the last two indices we can recast \eqref{bwc3aj} to the following form,
\be
J^{(\a_1...\a_{l-1}|\m|\a_l)\n}=\frac14\left[I^{(\a_1...\a_{l-1}\a_l)\m\n}-I^{(\m(\a_1...\a_{l-1})\a_l)\n}-I^{(\n(\a_1...\a_{l-1}\a_l)\m}+I^{(\m\n(\a_1...\a_{l-2})\a_{l-1}\a_l)}\right]\;.
\ee
We now use the constrain \eqref{gr7xews} and notice that
\be
I^{(\a_1...\a_{l-1}\a_l\m)\n}=\frac{1}{l+1}\left[I^{\a_1...\a_{l-1}\a_l\m\n}+lI^{(\m(\a_1...\a_{l-1})\a_l)\n}\right]=0\;,
\ee
which gives
\be\label{m4bvz}
I^{(\m(\a_1...\a_{l-1})\a_l)\n}=-\frac{1}{l} I^{\a_1...\a_{l-1}\a_l\m\n}\;,
\ee
and, because of the symmetry with respect to indices $\m$ and $\n$,
\be\label{h3fs5a}
I^{(\n(\a_1...\a_{l-1})\a_l)\m}=-\frac{1}{l} I^{\a_1...\a_{l-1}\a_l\m\n}\;.
\ee
We also have
\ba
I^{(\a_1...\a_{l-1}\a_l\m\n)}&=&\frac{2!l!}{(l+2)!}\\\nonumber
&\times&\left[I^{\a_1...\a_{l-1}\a_l\m\n}+l I^{(\m(\a_1...\a_{l-1})\a_l)\n}+l I^{(\n(\a_1...\a_{l-1})\a_l)\m}+\frac{l(l-1)}{2}I^{(\m\n(\a_1...\a_{l-2})\a_{l-1}\a_l)}\right]=0\;,
\ea
which yields
\be\label{jerc2}
I^{(\m\n(\a_1...\a_{l-2})\a_{l-1}\a_l)}=\frac{2}{l(l-1)}I^{\a_1...\a_{l-1}\a_l\m\n}\;.
\ee
Replacing \eqref{m4bvz}, \eqref{h3fs5a} and \eqref{jerc2} to \eqref{bwc3aj} yields
\be\label{x8d4ak} 
J^{(\a_1...\a_{l-1}|\m|\a_l)\n}=\frac14\frac{l+1}{l-1}I^{\a_1...\a_l\m\n}\;,
\ee
that shows the algebraic equivalence between the symmetrical $J^{(\a_1...\a_{l-1}|\m|\a_l)\n}$ and $I^{\a_1...\a_l\m\n}$ multipole moments for $l\ge 2$. Due to the orthogonality condition \eqref{kerxv34} we conclude that 
\be \label{juevz5}
{\mathfrak n}_{\a_1}I^{\a_1...\a_l\m\n}=0\;,
\ee
for the first $l$ indices of $I^{\a_1...\a_l\m\n}$. The number of these conditions is the same as the number of components of tensor $I^{\a_1...\a_{l-1}\m
\n}$ that is $N_3(l-1)=(l+2)(l+1)(l-2)$. It reduces the number of linearly independent components of $I^{\a_1...\a_l\m\n}$ to $N=N_3(l)-N_3(l-1)=(l+2)
(3l-1)$ \citep{dixon_1979,Dixon2015}. This exactly corresponds to the number of the linearly-independent components of tensor $J^{\a_1...a_{l-1}\m\a_l\n}$. 
Therefore, equation \eqref{wk12dcr} provides an easy way to compute the linearly-independent components of tensor $I^{\a_1...\a_l\m\n}$ which are the only components that matter in subsequent computations. 

\subsection{The Stress-Energy Skeleton and the Dixon Multipoles}

At this point of our discussion, we notice that the original definition \eqref{q12ttt} of multipoles $I^{\a_1...\a_l\m\n}$ contains the time components, $X^0$, of vector $X^\a$ which are nonphysical as they cannot be measured by a local observer with dynamic velocity ${\mathfrak n}^\a$ at point $z$ on the reference worldline ${\cal Z}$. Only those components of $I^{\a_1...\a_l\m\n}$ which are orthogonal to ${\mathfrak n}^\a$ can be measured. This explains the physical meaning of the orthogonality condition \eqref{juevz5}. Taking into account this observation, it is reasonable to introduce a new notation for the physically-meaningful components of Dixon's multipoles, 
\be\label{p3gz7}
{\cal J}^{\a_1...\a_l\m\n}=P^{\a_1}_{\b_1}...P^{\a_l}_{\b_l}\int_\Sigma X^{\b_1}...X^{\b_l}\hat T^{\m\n}(z,X)\sqrt{-\bar g(z)}d\Sigma\;,
\ee
where the integration is performed in 4-dimensional spacetime over the hypersurface $\Sigma$ passing through the point $z$ with the element of integration $d\Sigma={\mathfrak n}^\a d\Sigma_\a$, and 
\be 
P^{\a}_{\b}=\d^{\a}_{\b}+{\mathfrak n}^\a {\mathfrak n}_\b\;,
\ee
is the operator of projection on the hypersurface $\Sigma$ making all vectors $X^\a$ in \eqref{p3gz7} orthogonal to ${\mathfrak n}^\a$. The multipoles ${\cal J}^{\a_1...\a_l\m\n}$ have the same symmetries \eqref{o39v3j}, \eqref{gr7xews} as $I^{\a_1...\a_l\m\n}$, 
\ba \label{mev49}
{\cal J}^{\a_1...\a_l\m\n}&=&{\cal J}^{(\a_1...\a_l)(\m\n)}\;,\\\label{bstw5v}
{\cal J}^{(\a_1...\a_l\m)\n}&=&0\;,
\ea
while the orthogonality condition \eqref{juevz5} is identically satisfied and is no longer considered as an additional constraint. The projection operator is idempotent, that is obey the following rule
\be\label{kkk9}
P^\a_\g P^\g_\b=P^\a_\b\;,
\ee
which makes only 3 out of 4 components of $X^\a$ linearly-independent in \eqref{p3gz7}. On the other hand, the indices $\m$ and $\n$ in ${\cal J}^{\a_1...\a_l\m\n}$ still take values from the set $\{0,1,2,3\}$. Thus, equation \eqref{mev49} tells us that the overall number of components of ${\cal J}^{\a_1...\a_l\m\n}$ is $C^{l+2}_2\times C^5_3=5(l+2)(l+1)$ while the number of constraints \eqref{bstw5v} is $C^{l+2}_2\times C^4_3=2(l+3)(l+2)$. It gives the number of the algebraically-independent components of ${\cal J}^{\a_1...\a_l\m\n}$ equal to $N=(l+2)(3l-1)$ which exactly coincides with the number of algebraically-independent components of Dixon's multipoles $I^{\a_1...\a_l\m\n}$. 

Picking up the local Riemann coordinates in such a way that $X^0$ component of vector $X^\a$ is directed along the dynamic velocity ${\mathfrak n}^\a$ and three other components $X^i=\{X^1,X^2,X^3\}$ are lying in the hypersurface $\Sigma$, yields skeleton's structure, 
\be\label{hhrr6c3}
\hat T^{\m\n}(z,X)=\int_{-\infty}^{+\infty}\d(X^0)\hat T^{\m\n}_\perp(X^i)dX^0\;,
\ee
where $\d(X^0)$ is Dirac's delta-function and the distribution $\hat T^{\m\n}_\perp\in\Sigma$. Substituting \eqref{hhrr6c3} to \eqref{p3gz7} and taking into account that in these coordinates $DX=dX^0d\Sigma$, we obtain that Dixon's multipoles $I^{\a_1...\a_l\m\n}={\cal J}^{\a_1...\a_l\m\n}$ and, due to the tensor nature of the multipoles, this equality is retained in arbitrary coordinates.   

Exact nature of the distribution $\hat T^{\m\n}_\perp(X^i)$ in full general relativity is not yet known due to the non-linearity of the Einstein equations. Nonetheless, the Dirac delta-function is a reasonable candidate being sufficient to work in the post-Newtonian approximation with a corresponding regularization techniques \citep{blanchet_2001JMP}. For the purpose of the present paper it is sufficient to assume that in arbitrary coordinates the stress-energy skeleton \eqref{hhrr6c3} has the following structure \citep{Ohashi_2003PRD,steinhoff_2010PhRvD,Pound_2015} 
\be\label{bb22sz}
\hat T^{\m\n}(z,x)=\sum_{l=0}^\infty \int_{-\infty}^{+\infty}\bar\nabla_{\a_1...\a_l}\bigg[{\mathsf t}^{\a_1...\a_l\m\n}(z)\frac{\d_4\left(x-z\right)}{\sqrt{-\bar g(z)}}\bigg]\frac{ds}{\sqrt{-\bar g_{\m\n}(z) {\mathfrak n}^\m  {\mathfrak n}^\n}}\;,
\ee
where $s$ is an affine parameter along the geodesic in direction of the dynamic velocity ${\mathfrak n}^\a$, $\d_4\left(x-z\right)\equiv \d_4\left[x^\a-z^\a(s)\right]$ is 4-dimensional Dirac's delta-function, ${\mathsf t}^{\a_1...\a_l\m\n}$ are generalized multipole moments defined on the worldline ${\cal Z}$ that are orthogonal to ${\mathfrak n}^\a$ in the first $l$ indices (${\mathfrak n}_{\a_1}{\mathsf t}^{\a_1...\a_l\m\n}=0$), and $\bar\nabla_{\a_1...\a_l}\equiv \bar\nabla_{\a_1}...\bar\nabla_{\a_l}$ is a covariant derivative of the order $l$ taken with respect to the argument $x\equiv x^\a$ of the Dirac delta-function on the background manifold $\bar M$. Notice that expression \eqref{bb22sz} is a simplification of the original Mathisson theory \citep{mathisson_2010GReGr_1,mathisson_2010GReGr_1} proposed by \citet{tulczyjew1}.  \citet{dixon_1979} did not specify the nature of the singularity entering definition \eqref{bb22sz} assuming that Dirac's delta-function is solely valid in the pole-dipole approximation while a more general type of distribution is required in the definition of the stress-energy {skeleton} for high-order multipoles. The Dirac delta-function is widely adopted in computations of equations of motion of relativistic binary systems \citep{schaefer_2011mmgr,spin_Hamiltonian_schaefer,Blanchet_2002LRR} amended with corresponding regularization techniques to deal with the singularities in the non-linear approximations of general relativity \citep{Damour_1987book,blanchet_2004PhRvD,blanchet_2005PhRvD7,Dixon_2013}.

The generalized multipoles ${\mathsf t}^{\a_1...\a_l\m\n}$ are used to derive the explicit form of the MPD equations of motion in terms of the linear momentum ${\mathfrak p}^\a$, angular momentum $S^{\a\b}$ and Dixon's multipole moments $I^{\a_1...\a_l\m\n}$ as demonstrated by \citet{mathisson_2010GReGr_1,mathisson_2010GReGr_2}, \citet{pap1,pap2}, \citet{dixon_1979} and other researchers   \citep{bini_2009GReGr,dirk_obukhov_2015PhRvD,steinhoff_2010PhRvD,dirk_obukhov2014,Ohashi_2003PRD}.  It turns out that the generalized multipoles ${\mathsf t}^{\a_1...\a_l\m\n}$ are effectively equivalent to the body multipoles, ${\cal J}^{\a_1...\a_l\m\n}$. Indeed, replacing the stress-energy skeleton \eqref{bb22sz} to \eqref{q12ttt}, transforming the most general coordinates $x^\a$ in \eqref{bb22sz} to the local Riemannian coordinates $X^\a$,  and taking the covariant derivatives yield 
\be\label{ncv4d}
{\cal J}^{\a_1...\a_l\m\n}=P^{\a_1}_{\b_1}...P^{\a_l}_{\b_l}\sum_{n=0}^\infty {\mathsf t}^{\g_1...\g_p\m\n}\int X^{\b_1}...X^{\b_l}\frac{\pd^n \d_4(X)}{\pd X^{\g_1}...\pd X^{\g_n}}DX\;.
\ee
Integrating by parts, taking the partial derivatives from $X^\a$, and accounting for the integral properties of delta-function \citep{shilov_1968}, we conclude 
\be\label{jj239}
{\cal J}^{\a_1...\a_l\m\n}=(-1)^ll!{\mathsf t}^{\a_1...\a_l\m\n}\;.
\ee
Relation \eqref{jj239} establishes connection between the Dixon multipoles and the generalized moments of the stress-energy skeleton.
 
\subsection{The Equivalence of the Dixon Multipoles and the Blanchet-Damour Multipoles}

To proceed further on, we shall assume that the dynamic velocity ${\mathfrak n}^\a$ is equal to the kinematic velocity ${u}^\a$. This assumption is consistent with Dixon's mathematical development and agrees with our covariant definition \eqref{zowv34as} of the linear momentum of an extended body moving on the background spacetime manifold $\bar M$. It also allows us to employ
the results obtained previously by \citet{Ohashi_2003PRD}, to retrieve  a covariant expression for the generalized multipoles ${\mathsf t}^{\a_1...\a_l\m\n}$ of the gravitational skeleton $\hat T^{\m\n}$ from the multipolar expansion of the metric tensor of a single body. We assume in this approach that the skeleton \eqref{bb22sz} is the source of the gravitational field in the Einstein field equations for the metric tensor. This assumption was never made explicit in Dixon's papers \citep{dixon_1979,Dixon2015} because the field equations are circumvented in the Mathisson variational dynamics. Nonetheless, in general relativity the multipole moments of gravitational field must be intimately related to the source of gravity. Thus, it is natural to assume that the stress-energy skeleton $\hat T^{\a\b}$ entering the definition of the Dixon multipole moments is identical to the stress-energy skeleton which is a source of gravitational field generated by matter of extended bodies in $N$-body system.

We derive the generalized multipoles of the stress-energy skeleton of body B from \citep[Equation 3.1]{Ohashi_2003PRD} after reconciling the sign conventions for the metric tensor perturbation and for the normalization coefficients of multipoles adopted in \citep{Ohashi_2003PRD} with those adopted by \citet[Equation 2.32]{bld1986} which we use everywhere in the present paper. The generalized moments of the stress-energy {skeleton} read,  
\ba\label{gg55zz99} 
{\mathsf t}^{\a_1...\a_l\m\n}&=&\frac{(-1)^l}{l!}\left[{u}^\m {u}^\n{\cal M}^{\a_1...\a_l}+\frac{2}{l+1}{u}^{(\m}\dot{\cal M}^{\n)\a_1...\a_l}+\frac{1}{(l+1)(l+2)}\ddot{\cal M}^{\m\n\a_1...\a_l}\right]\\\nonumber
&-&\frac{(-1)^l}{l!}\left[\frac{2l}{l+1}{u}^{(\m}{\varepsilon}_\b{}^{\n)<\a_1}{\cal S}^{\a_2...\a_l>\b}+\frac{2}{l+2}\varepsilon_\b{}^{<\a_1(\m}\dot{\cal S}^{\n)\a_2...\a_l>\b}\right]\;,
\ea
where the dot above functions denotes the Fermi-Walker covariant derivative \eqref{p2b8r3} and \eqref{huc4} and all multipoles are purely spatial Cartesian STF tensors in the sense of orthogonality to 4-velocity $u^\mu$ as shown in \eqref{m1zz1}.
Comparing \eqref{gg55zz99} with \eqref{jj239} we obtain the relationship between the Dixon internal multipoles and the Blanchet-Damour mass and spin multipoles used in the present paper,
\ba\label{ney1c} 
{\cal J}^{\a_1...\a_l\m\n}&=&{u}^\m {u}^\n{\cal M}^{\a_1...\a_l}+\frac{2}{l+1}{u}^{(\m}\dot{\cal M}^{\n)\a_1...\a_l}+\frac{1}{(l+1)(l+2)}\ddot{\cal M}^{\m\n\a_1...\a_l}\\\nonumber
&-&\frac{2l}{l+1}{u}^{(\m}{\varepsilon}_\b{}^{\n)<\a_1}{\cal S}^{\a_2...\a_l>\b}-\frac{2}{l+2}\varepsilon_\b{}^{<\a_1(\m}\dot{\cal S}^{\n)\a_2...\a_l>\b}\;,
\ea
We still have to take into account the identity \eqref{bstw5v} in order to eliminate linearly-dependent components of ${\cal J}^{\a_1...\a_l\m\n}$. The most easy way is to account for it is to take the double skew-symmetric part with respect to the last four indices as shown in the right hand side of equation \eqref{wk12dcr}. It yields 
\be\label{cbxczw}
I^{\a_1...\a_l\m\n}\equiv {\cal J}^{\a_1...[a_{l-1}[\a_l\m]\n]}=4\left\{{\cal M}^{\a_1...[\a_{l-1}[\a_l}{ u}^{\m]}{u}^{\n]}+\frac{l}{l+1}{\cal S}^{\b<\a_1...[\a_{l-1}}{ u}^{(\m]}{\varepsilon}^{\a_l>\n)}{}_\b\right\}\;,
\ee
where we have taken into account that in calculating the skew-symmetric part of 4-velocity $u^\mu$ with a purely spatial tensor we have, for example, 
\be
{\cal M}^{\a_1...\a_{l-1}[\a_l}{ u}^{\m]}=\pi^{\a_l}_{\b_l}{\cal M}^{\a_1...\a_{l-1}[\b_l}{\bar u}^{\m]}=\frac12{\cal M}^{\a_1...\a_{l-1}\a_l}{ u}^{\m}\;,
\ee
and so on.

The relation between Dixon's $J$ and $I$ multipole moments has been deduced in \eqref{x8d4ak}. Substituting expression \eqref{cbxczw} for the Dixon multipoles $I$ in the right hand side of \eqref{x8d4ak} provides a correspondence between the symmetrized Dixon multipoles $J$ and the Blanchet-Damour mass and spin multipoles in the following form 
\ba\label{ut3cw}
J^{(\a_1...\a_{l-1}|\m|\a_l)\n}&=&\frac{l+1}{l-1}\left\{{\cal M}^{\a_1...[\a_{l-1}[\a_l}{u}^{\m]}{u}^{\n]}+\frac{l}{l+1}{\cal S}^{\b<\a_1...[\a_{l-1}}{u}^{(\m]}{\varepsilon}^{\a_l>\n)}{}_\b\right\}\;,
\ea
where the vertical bars around index $\mu$ in the left hand side of \eqref{ut3cw} means that this index is excluded from the symmetrization.
Equation \eqref{ut3cw} demonstrates the total equivalence (up to the numerical factor $(l+1)/(l-1)$) of the Dixon and Blanchet-Damour multipoles.

\section{Post-Newtonian Covariant Equations of Motion Versus the Dixon Equations of Motion}\label{appendixB}
\subsection{Comparison of translational equations for linear momentum}\label{ndj45xc}
In order to compare our translational equations of motion \eqref{ubpu;b} with Dixon's equation \eqref{q15ms} we need to symmetrize the covariant derivatives in the right hand side of \eqref{q15ms}. It is achieved with the help of the following algebraic transformation, 
\ba\label{yu564x}
\bar\nabla_{\a(\b_1...\b_{l-2}}R_{|\m|\b_{l-1}\b_l)\n} J^{\b_1...\b_{l-1}\m\b_l\n}&=&\bar\nabla_{(\a\b_1...\b_{l-2}}R_{|\m|\b_{l-1}\b_l)\n} J^{\b_1...\b_{l-1}\m\b_l\n}\\\nonumber
&+&
\frac{2}{l+1}\bar\nabla_{\n(\b_1...\b_{l-2}}R_{|\m|\b_{l-1}\b_l)\a} J^{\b_1...\b_{l-1}\m\b_l\n}+{\cal O}(R^2)\;,
\ea
where the residual terms are proportional to the square of the Riemann tensor, and have been discarded. These quadratic-in-curvature terms are important for the post-Newtonian equations of motion but complicate the equations which follow and, hence, will be omitted every time when they appear. Substituting \eqref{ut3cw} to the right hand side of \eqref{yu564x} yields
\ba\label{bye62c}
\bar\nabla_{\a(\b_1...\b_{l-2}}R_{|\m|\b_{l-1}\b_l)\n} J^{\b_1...\b_{l-1}\m\b_l\n}&=&\frac{l+1}{l-1}\left[{\cal E}_{\a\b_1...\b_l}{\cal M}^{\b_1...\b_l}+\frac{l}{l+1}{\cal C}_{\a\b_1...\b_l}{\cal S}^{\b_1...\b_l} \right]+{\cal O}(R^2)\;,
\ea
where the external multipole moments ${\cal E}_{\a_1...\a_l}$ and ${\cal C}_{\a_1...\a_l}$ have been defined in \eqref{sd11} and \eqref{we70} respectively.
Computation of the second term in the right hand side of \eqref{yu564x} shows that it is of the second order in curvature tensor, and can be omitted as we have agreed above.

Substituting \eqref{bye62c} to the right hand side of \eqref{q15ms} recasts it to
\ba\label{hwcz41}
\frac{{\cal D} {p}_\a}{{\cal D}\tau}&=&\frac12 {u}^\b S^{\m\n}\bar R_{\m\n\b\a}+\sum\limits_{l=2}^{\infty}\frac{1}{l!} \left[{\cal E}_{\a\b_1...\b_l}{\cal M}^{\b_1...\b_l}+\frac{l}{l+1}{\cal C}_{\a\b_1...\b_l}{\cal S}^{\b_1...\b_l} \right]+{\cal O}(R^2)\;,
\ea
The very first term in the right hand side depending on $S^{\a\b}$,  can be incorporated to the sum over the spin moments by making use of the duality relation between body's intrinsic spin ${\cal S}^\a$ and spin-tensor $S^{\a\b}$ \footnote{The minus sign in \eqref{ju3v8m} appears because Dixon's definition \eqref{wk12} of $S^{\a\b}$ has an opposite sign as compared to our definition \eqref{spin-9} of spin. }
\be\label{ju3v8m}
 S^{\m\n}=-\varepsilon^{\m\n}{}_\a{\cal S}^\a\;,
 \ee
 where the Levi-Civita tensor $\varepsilon_{\a\b\g}$ has been defined above in \eqref{vareps67}. It yields
 \be  
 {u}^\b S^{\m\n}\bar R_{\m\n\b\a}={\cal C}_{\a\b}{\cal S}^\b\;,
 \ee
where ${\cal C}_{\a\b}$ is given by \eqref{we70} for $l=2$. Making use of \eqref{ju3v8m}  allows to rewrite \eqref{hwcz41} in the final form
\be\label{uvnet467}
\frac{{\cal D} {p}_\a}{{\cal D}\tau}=\sum\limits_{l=2}^{\infty}\frac{1}{l!} {\cal E}_{\a\b_1...\b_l}{\cal M}^{\b_1...\b_l}+\sum\limits_{l=1}^{\infty} \frac{l}{(l+1)!}{\cal C}_{\a\b_1...\b_l}{\cal S}^{\b_1...\b_l}+{\cal O}(R^2)\;.
\ee

Thus, Dixon's equation of translational motion \eqref{q15ms} given in terms of Dixon's internal multipoles and Veblen's tensor extensions of the Riemann tensor are brought to the form \eqref{uvnet467} given in terms of the gravitoelectric, ${\cal E}_{\a_1...\a_l}$, and gravitomagnetic, ${\cal C}_{\a_1...\a_l}$, external multipoles as well as mass, ${\cal M}^{\a_1...\a_l}$ and spin, ${\cal S}^{\a_1...\a_l}$ Blanchet-Damour internal multipoles. Comparing with the post-Newtonian covariant form of the translational equations of motion \eqref{ubpu;b}--\eqref{sd16} one can see that the right hand side of the Dixon equation \eqref{uvnet467} reproduces only two terms in the covariant expression for the post-Newtonian force \eqref{g6xc2c7}--\eqref{sd16}, more specifically -- the very first term of force $F^\a_{\cal Q}$ in \eqref{sd15a} and that of $F^\a_{\cal C}$ in \eqref{sd16}. The terms which are absent in the right hand side of Dixon's translational equations of motion \eqref{uvnet467} but are present in our equations \eqref{ubpu;b}--\eqref{sd16} include the $\dot{\cal M}$ term in \eqref{ubpu;b}, the quadratic-in-curvature terms shown in \eqref{we69} and the terms, which depend on the time derivatives of the STF multipoles both external and internal ones. 

The terms with the time derivatives of the multipoles are expected to be present in the equations of motion in the most general case because they reflect the temporal changes in the distribution of matter density and matter current inside body B as well as changes of the tidal gravitational field along the worldline ${\cal Z}$ of the center of mass of body B. We notice that some terms with the time derivatives of the multipoles in the right hand side of our equation \eqref{ubpu;b} can be grouped together to form the total time derivative which can be included to definition \eqref{zowv34as} of linear momentum $\mathfrak{p}^a$. This allows to eliminate any linear combination of terms from the right hand side of \eqref{ubpu;b} which form the total time derivative. However, not all terms in the right hand side of \eqref{ubpu;b} are reduced to the total time derivatives so we cannot reach a complete equivalence between equation \eqref{ubpu;b} and \eqref{uvnet467} by redefining mass, ${\cal M}=m+\Delta m$, and linear momentum, $\mathfrak{p}^a=p^\a+\Delta p^\a$, with some properly adjusted scalar and vector functions $\Delta m$ and $\Delta p^\a$. The difference between the two forms of the post-Newtonian covariant equations -- \eqref{ubpu;b} and \eqref{uvnet467}, has a more principal character and its origin is not yet clear. It may be due to the choice of the gauge conditions imposed on the metric tensor. We have derived our covariant equations \eqref{ubpu;b}--\eqref{sd16} by imposing the harmonic gauge \eqref{hhb3vx5z} while derivation of the Dixon equations of motion \eqref{q15ms} is based on the choice of the RNC gauge \eqref{rnc_gauge} on the effective background manifold $\bar M$. 

The possible appearance of the time derivatives of the multipole moments in the post-Newtonian equations of motion has been never pointed out in Dixon's papers \citep{dixon_1979,dixon_2008} nor in the papers of other researchers exploring the other various aspects of the MPD formalism. More work is required to take into account all possible contributions to the post-Newtonian covariant translational equations of motion \eqref{uvnet467} derived in the framework of the Mathisson variational dynamics. Constructive steps towards further developing the MPD formalism are discussed in review papers by Dixon \citep{Dixon2015} and Harte \citep{harte2015}.     

\subsection{Comparison of the rotational equations for spin}\label{iopn3e4}

Dixon's equations of rotational motion for spin are given by equation \eqref{q16mc}. We express spin ${\cal S}^\a$ of body B in terms of the spin tensor $S^{\lambda\sigma}$ by inverting \eqref{ju3v8m},
\be \label{lec27b}
{\cal S}^\a=-\frac12\varepsilon^{\a\lambda\s}S_{\lambda\s}\;.
\ee 
We take covariant derivative from both side of \eqref{lec27b} with accounting for a covariant derivative from 3-dimensional Levi-Civita tensor defined in \eqref{vareps67} 
\be\label{jk3c9}
\frac{{\cal D}\varepsilon^{\a\lambda\s}}{{\cal D}\tau}=\frac{1}{\sqrt{-\bar g}}{\cal Q}_\m E^{\m\a\lambda\s}\;,
\ee
and replacing the covariant derivative from $S^{\lambda\s}$ with the terms from the right hand side of \eqref{q16mc}. It yields,
\be \la{op4vx3ww}
\frac{{\cal D}{\cal S}^\a}{{\cal D}\tau}=-\left({\cal S}^\b{\cal Q}_\b\right)u^\a-\varepsilon^\a{}_{\lambda\s}\sum\limits_{l=1}^{\infty}\frac{1}{l!}\bar\nabla_{(\b_1...\b_{l-1}}\bar R_{|\m|\r\b_l)\n}g^{\r\lambda}
\left[{\cal M}^{\s\b_1...\b_{l-1}\b_l}{u}^{\m}{u}^{\n}+\frac{l+1}{l+2}{\cal S}^{\s\g\b_1...\b_{l-1}}{u}^{\m}{\varepsilon}^{\b_l\n}{}_\g\right]\;,
\ee
where we have also used \eqref{ut3cw} to replace the Dixon internal multipole moments with the Blanchet-Damour mass and spin multipoles \footnote{Notice that the term $\varepsilon^\a{}_{\lambda\s}u^{[\lambda} p^{\s]}=0$ due to the orthogonality of $p^\sigma$ and $u^\lambda$ to $\varepsilon^\a{}_{\lambda\s}$.}. Now, we employ the covariant definitions \eqref{sd11} and \eqref{we70} of the gravitoelectric and gravitomagnetic external multipoles in \eqref{op4vx3ww} which takes on the following form,
\be\label{xcz5ds}
\frac{{\cal D}{\cal S}^\a}{{\cal D}\tau}=-\left({\cal S}^\b{\cal Q}_\b\right)u^\a-\varepsilon^{\a\lambda}{}_\s\sum\limits_{l=1}^{\infty}\frac{1}{l!}
\left[{\cal E}_{\lambda\b_1...\b_l}{\cal M}^{\s\b_1...\b_l}+\frac{l+1}{l+2}{\cal C}_{\lambda\b_1...\b_l}{\cal S}^{\s\b_1...\b_l}\right]\;.
\ee
Now, we can compare Dixon's equation of rotational motion \eqref{xcz5ds} for spin of body B with our equations \eqref{ui6s41}-\eqref{ac5s03v}. As we can see they are in perfect agreement. A small difference between \eqref{xcz5ds} and \eqref{ui6s41} is in the quadratic-in-curvature terms which we have taken into account in the torque \eqref{ac5s03v}. 

\section{Acknowledgments}
I am thankful to W.~G. Dixon, A.~I. Harte and A.~A. Deriglazov for valuable conversations and to an anonymous referee for constructive report that has helped to improve presentation of the article.  
\newpage

\bibliographystyle{unsrtnat}
\bibliography{Kopeikin_focus_point_article}
\end{document}